\documentclass[preprint,aps]{revtex4}
\usepackage{color}
\usepackage{xcolor}
\usepackage{mathtools}
\usepackage{diagbox}
\usepackage{appendix}
\usepackage{amsmath,amssymb,amsfonts,dcolumn,color,graphicx,graphics,latexsym,epsfig}
\usepackage[compatibility=false]{caption}
\usepackage{subcaption}
\usepackage{hyperref}
\usepackage{natbib}
\usepackage{float}
\usepackage{booktabs}
\def\beq{\begin{equation}}
\def\eeq{\end{equation}}
\def\barr{\begin{array}}
\def\earr{\end{array}}

\begin{document}

\title{Revisiting a family of wormholes: geometry, matter, scalar quasinormal modes
and echoes}

\author{Poulami Dutta Roy${}^\#$, S. Aneesh${}^{\dagger}$ and
Sayan Kar ${}^{\#, *}$}
\email{poulamiphysics@iitkgp.ac.in, aneeshs@ufl.edu, sayan@phy.iitkgp.ac.in}
\affiliation{${}^\#$Department of Physics, Indian Institute of Technology 
Kharagpur, 721 302, India}
\affiliation{${}^{\dagger}$ 
Department of Physics, 2001 Museum Road, P.O. Box 118440, University of Florida,
Gainesville, FL 32611-8440}
\affiliation{${}^{*}$
Centre for Theoretical Studies, Indian Institute of Technology, Kharagpur, 
721 302, India}

\begin{abstract}
\noindent  
We revisit a family of ultra-static Lorentzian wormholes which includes
Ellis-Bronnikov spacetime as a special case. We first show how the
required total matter stress energy (which violates the local energy conditions) may be split 
into a part due to a phantom scalar and
another extra piece (which vanishes for Ellis--Bronnikov) satisfying the Averaged Null Energy Condition (ANEC) along radial null geodesics. Thereafter,
we examine 
the effective potential for scalar wave propagation in a general setting.  Conditions 
on the metric function, for which the effective potential may have double barrier features are written down and illustrated (using this class of wormholes).  
Subsequently, using numerous methods, we obtain the scalar quasinormal modes (QNMs). We note the behaviour 
of the QNMs as a function of $n$ (the metric parameter) and $b_0$ (the wormhole throat
radius).
Thus, the 
shapes and sizes of the wormholes, governed by the metric parameter $n$ and the throat radius
$b_0$ are linked to the variation and the values of the
QNMs. Finally, we demonstrate how, for large $n$, the time domain
profiles exhibit, expectedly, the occurence of echoes. In summary, our results suggest that this family of wormholes
may indeed be used as a template for further studies on the gravitational wave physics of exotic compact objects. 
\end{abstract}


\maketitle

\newpage

\section {\bf Introduction}

\noindent Much of the interest today in traversable Lorentzian wormhole spacetimes (originally proposed in \cite{einstein_1,einstein_2, wheeler})
revolve around the question: {\em do they exist?} While the existence of black holes
is no longer in doubt (more so after recent observations in M87 \cite{m87}), the
wormhole story is far from complete. The same may be said about naked singularities
and cosmic censorship too \cite{naked_1,naked_2}. 

\noindent The existence question on wormholes is based on a couple of issues. Firstly,
classical General Relativity (GR) along with the imposed energy conditions does not allow wormholes \cite{mt,mty,ec11, ec12,ec13}.
In other words,
the shape of the spatial slice of a wormhole is such that a converging null geodesic
congruence would have to be defocused, as long as a throat and a `flare out to
the other universe' (second asymptotically flat region)-- both necessary
geometric features of wormholes--have to be admitted \cite{wormholes,hawk,wald}. 
To get away with this so-called `defect' or `problem'
one may appeal to modified theories of gravity. In such theories, the energy conditions 
on matter may hold but the convergence condition is violated \cite{rajibul}, \cite{modified_quadratic}. Examples of 
wormholes in modified gravity are numerous \cite{alt_1,alt_2,alt_3} and they largely have
energy-condition-satisfying matter \cite{wormmod_Weyl_1,wormmod_Weyl_2, wormmod_Lovelock, wormmod_bumblebee, wormmod_f(R), wormmod_f(R)1, wormmod_f(T), wormmod_STT, wormmod_ECT,wormmod_EsGB_1,wormmod_EsGB_2, wormmod_DEsGB_1,wormmod_DEsGB_2, wormmod_EBI, wormmod_scalar} and sometimes, non-phantom fields \cite{ECT}, \cite{EsGB}. In addition to taking refuge
in modified gravity, other ways of {\em restricting} the violation of
energy conditions are known. These include dynamic wormholes \cite{sk1,sk2,sk3,sk4,sk5}, a proposal for limiting the
amount of exotic matter \cite{ec11}  etc.

\noindent The second issue concerns possible signatures. Till recently, the most
compelling suggested signature for wormholes appeared to be from gravitational lensing
in such spacetimes \cite{lensing_1,lensing_2,lensing_3}.
However, with the advent of gravitational wave astronomy \cite{gwastro_1,gwastro_2,gwastro_3,gwastro_4,gwastro_5,gwastro_6}, one 
comes across the notion of {\em black hole mimickers} \cite{bhm1,bhm2,bhm3, bhm_WH_1,bhm_WH_2, bhm_WH_3, bhm_boson_1,bhm_boson_2,bhm_boson_3, bhm_gravastar_1,bhm_gravastar_2,bhm_gravastar_3} (eg. wormholes, gravastars
and other ultra-compact objects)
which can ideally mimic the results found using black holes, in GW observations. 
In other words, one may, in some 
scenarios, be able to explain GW observations using such black hole mimickers as
the end state of black hole  and/or neutron star mergers \cite{mergers_1,mergers_2}. Therefore, to improve upon
viable templates for black hole mimickers, it is necessary to study various properties
associated with them. One such property is a study of the quasinormal modes which can
be used by observers at GW interferometers for verifying a wormhole proposal. 

\noindent It is therefore of importance to study quasinormal modes of different types of
wormhole spacetimes. This has been done to some extent in \cite{bhm_WH_2,konoplya_zhidenko_2, konoplya, kerr_WH, kim, konoplya_molina, kokkotas, aneesh, kunz}. One of the purposes of this article is to further this line of thought
for another class of ultra-static wormholes. 

\noindent The choice of the family of wormholes we make here is based on an
earlier paper \cite{sksnm} where the well-known Ellis-Bronnikov spacetime \cite{ellis_1,ellis_2,ellis_3,ellis_4}
had been extended to provide a two-parameter family of spacetimes. We first show how
this generalisation leads to a matter-stress energy which has (a) an energy condition
violating part generated via a phantom scalar field and (b) another piece (of zero
value for Ellis-Bronnikov) which satisfies
the Averaged Null Energy Condition (ANEC) along radial null geodesics. 

\noindent The scalar field which generates the wormhole can therefore be perturbed
, with the ensuing perturbations satisfying $\Box \phi =0$. 
Thus, scalar quasinormal modes exist and are of interest in the context of this
family of wormholes. Surely, in future, gravitational perturbations
will have to be looked at, in order to make direct contact with GW observations. One may also recall that scalar
perturbations do appear to be relevant in modified theories of gravity, where
they arise as a so-called extra `breathing mode', which is absent in
GR. Numerous ways of detecting such a mode using a network of detectors have
been discussed in the literature \cite{breathing,scalar}. 

\noindent 
Several types of wormhole spacetimes have been investigated recently, which may be
thought of as {\em templates} for future studies especially with reference to GW observations. For most of these spacetimes, there
is no clear understanding of the matter required, from the viewpoint of a Lagrangian based
field theory coupled to gravity. One such 
well--known example is Damour-Solodukhin (DS) spacetime \cite{bhm_WH_1,kerr_WH,kokkotas,DS1,DS2,DS3,DS4}. Another interesting geometry is the black-bounce metric constructed in \cite{bounce} where a continuous parameter in the metric allows a transition from a regular BH to a one-way wormhole with an extremal null throat (black bounce) and then to a traversable wormhole.
The spacetime we work with here has distinctive features and may be viewed as another such
wormhole template.

\noindent 
Our work reported in this article is organised as follows. In
Section II, we introduce the wormhole spacetime.
We write down the Einstein tensor and equate it to the energy-momentum tensor of the `required matter'. We also show how the required matter can be modeled with a phantom scalar and
another extra piece satisfying ANEC.  The embedding diagram and the behaviour of the expansion
of a geodesic congruence is also briefly discussed.
In Section III, we move on to studying scalar wave propagation. We show when the
$s$-wave effective potential can behave as a double barrier by deriving the required
conditions on the metric functions.
Section IV is devoted to the scalar quasinormal modes in this class of spacetimes
The interesting possibility 
of the QNMs being used as a tool to determine the shape of a wormhole geometry in this family is presented here too. Further, the occurence of echoes in the time domain profile is
shown and analysed in
Section V.
Finally, in Section VI we end 
with  our conclusions and remarks. 

\section {\bf The family of ultra-static spacetimes} 
\noindent In their 1973 papers \cite{ellis_1,ellis_2}, Ellis and Bronnikov, independently constructed a spacetime using a phantom  (negative kinetic energy) scalar field source. Their work produced a static, spherically symmetric, geodesically complete, horizonless manifold with a throat (which Ellis called a `drainhole') connecting two asymptotically flat regions. The line element of the spacetime constructed by Ellis and Bronnikov is given as,
\begin{align}
    ds^{2} = -dt^{2} + \frac{dr^{2}}{1-\frac{b_{0}^{2}}{r^{2}}}+ r^{2} d\theta^{2} + r^{2} sin^{2}\theta d\phi^{2}.
    \label{eq:ellis_metric}
\end{align}
where $b_{0}$ is the throat radius of the `drainhole'. This spacetime is known today as the Ellis-Bronnikov wormhole. If the Morris-Thorne conditions \cite{mt} necessary for the construction of a Lorentzian wormhole are considered then one finds that there is ample scope for various similar wormhole geometries to exist. Thus, a generalised version of the Ellis-Bronnikov wormhole geometry was suggested in  \cite{sksnm} as a two-parameter ($n$ and the throat radius $b_0$) family of Lorentzian wormholes. When the parameter $n$ takes the value $n=2$, we get back the Ellis--Bronnikov spacetime. The motivation behind such a construction was to study the geodesics and propagation of scalar fields for a wider class of wormhole spacetimes and note various differences as well as similarities. The work also included the observation of resonances in the transmission coefficient for $n>2$ geometries thus indicating that $n>2$ geometries are clearly different from the Ellis-Bronnikov geometry ($n=2$). The line element of the generalised Ellis-Bronnikov spacetime involving the parameters $n$  and $b_0$ is given as,
\begin{align}
    ds^{2} = -dt^{2} + d\ell^{2}+ r^{2}(\ell) d\theta^{2} + r^{2}(\ell) sin^{2}\theta d\phi^{2} 
    \label{eq:gen_ellis_metric}
\end{align}
where 
\begin{align}
    r(\ell) = (\ell^{n}+b_{0}^{n})^{1/n}.
    \label{eq:radial_coord}
\end{align}
The parameter $n$ is allowed to take only even values to ensure the smooth behavior of $r(\ell)$ over the entire domain of the `tortoise' or `proper radial distance' coordinate $\ell$ ($- \infty \leq \ell \leq \infty)$. Note that the functional form of $r(\ell)$ has some curious features. At $\ell=0$ (wormhole throat), only the $n$-th derivative of $r(\ell)$ is non-vanishing. Also, the function $V(\ell) = \frac{r''}{r}$ (V($\ell$) is the effective potential discussed in Sec.III A) has a nonzero $n$-th derivative
at $\ell=0$, which is negative in value for $n=2$, but positive for all $n>2$. These facts will be crucial while discussing the effective potential for scalar wave propagation later in this article. 

\noindent The line element can also be written in an alternative form in terms of the usual radial coordinate $r$ as,
\begin{align}
     ds^{2} = -dt^{2} + \frac{dr^{2}}{1-\frac{b(r)}{r}}+ r^{2} d\theta^{2} + r^{2} sin^{2}\theta d\phi^{2}
\end{align}
where $r$ and $\ell$ are related through
\begin{gather}
    d\ell^2= \frac{dr^2}{1-\frac{b(r)}{r}}\\
    \Rightarrow b(r)=r-r^{(3-2n)} (r^n-b_0^n)^{(2-\frac{2}{n})}.
    \label{eq:shape_function}
\end{gather}

\subsection{\bf Geometry}
\noindent Given the metric functions in eqn.(\ref{eq:gen_ellis_metric}), we note that 
the spacetime is spherically symmetric and ultrastatic. All the metric components are independent of time 
and hence, all $t=constant$ slices are identical. This  will be useful when we embed a 2-D slice of the  
wormhole in flat space, in order to understand its shape. The embedding diagram encodes the shape of the wormhole through a variation of the shape function $b(r)$ or $r(\ell)$ (w.r.t the `$\ell$' coordinate). It is easy to observe that the $R_{00}$ component of the Ricci tensor will always be zero for this family of wormholes, irrespective of the form of $r(\ell)$ or $b(r)$. The spacetime geometry owes such features since the metric is ultra-static. We will see the distinction between geometries for different $n$ values explicitly when we plot the embedding diagrams for different $n$. One of the objectives of this paper is to distinguish the different wormhole geometries for different values of $n$ using the 
corresponding scalar quasi-normal modes.

\subsection{\bf Matter, energy conditions}
\noindent The energy conditions are a way of ensuring that a solution of Einstein's equations is physically viable. The creation and maintenance of any traversable wormhole was first studied by Morris and Thorne \cite{mt} through the energy conditions that they are supposed to satisfy. It was observed that for a traversable wormhole to exist in GR the Weak Energy Condition (WEC) must be violated
atleast at the throat.
This meant that one requires  {\em exotic matter} (i.e. matter violating the energy conditions). Later studies showed that all classes of static wormholes in GR need exotic matter for stability \cite{sk4,sk5,ec_visser}. This is a major drawback for wormholes and, as stated before, forbids their
existence within the tenets of GR.

\noindent Let us first write down the energy momentum tensor for a general $r(\ell)$ using the
Einstein equations and the Einstein tensor. The energy-momentum tensor defined by its diagonal components in the frame basis, i.e.  $T_{00}= \rho(\ell)$, $T_{11}= \tau(\ell)$, $T_{22}= T_{33}= p(\ell)$ has the following form with ($8\pi G=c^2=1$),
\begin{eqnarray}
\rho (\ell) = -2 \frac{r''}{r} - \left (\frac{r'}{r}\right )^2+\frac{1}{r^2}\\
\tau (\ell) = \left (\frac{r'}{r}\right )^2 - \frac{1}{r^2} \\
p(\ell) = \frac{r''}{r}
\end{eqnarray}
\noindent It is clear from the above expressions that, for any $r(\ell)$, $\rho+\tau=-2p=-2 \frac{r''}{r}$. 
We will see later, how the single or double barrier character of the effective potential
is related to the energy conditions and the nature of $r(\ell)$.
In general for any wormhole, $r(\ell)$ is preferably an even function, $r(0)= b_0$, and $r\sim \ell$ as $\ell\rightarrow \pm \infty$. Thus $r(\ell)$ always has a minimum at $\ell=0$.

\noindent We now write down the energy-momentum tensor for our wormhole spacetime.  
Similar to the case of general $r(\ell)$, we calculate $\rho$, $\tau$ and $p$ for our wormhole family using $x=\frac{\ell}{b_0}$, 
\begin{gather}
    \rho(x) = \rho_\phi + \rho_e = -\frac{1}{b_0^2} \left [ \frac{1}{(1+x^n)^{\frac{4}{n}}} \right ] +\frac{1}{b_0^2} \left [ \frac{1}{(1+x^{n})^{\frac{2}{n}}} +\frac{1}{(1+x^n)^{\frac{4}{n}}} -
    \frac{x^{2n-2} + 2 (n-1) x^{n-2}}{(1+x^{n})^2} \right ] \\
    \tau(x)=  \tau_\phi +\tau_e= -\frac{1}{b_0^2} \left [ \frac{1}{(1+x^n)^{\frac{4}{n}}} \right ] + 
    \frac{1}{b_0^2} \left [ \frac{x^{2n-2}}{(1+x^{n})^{2}}-\frac{1}{(1+x^{n})^{\frac{2}{n}}} + \frac{1}{(1+x^n)^{\frac{4}{n}}} \right ]\\
    p(x) = p_\phi + p_e =  \frac{1}{b_0^2} \left  [ \frac{1}{(1+x^n)^{\frac{4}{n}}} \right ]
+ \frac{1}{b_0^2} \left [ \frac{(n-1)x^{n-2}}{(1+x^{n})^{2}} - \frac{1}{(1+x^n)^{\frac{4}{n}}} \right ]
\end{gather}
We have written the $\rho,\tau,p$ as a sum of (i) a contribution from a phantom scalar field ($\rho_\phi, \tau_\phi, p_\phi$, i.e. the first terms in the R. H.S. of (10), (11), (12)) and (ii) extra matter
($\rho_e,\tau_e, p_e$), i.e. the second terms in the R. H. S. of (10), (11), (12). Recall that for $n=2$, the well--known Ellis--Bronnikov spacetime
can be obtained as an exact solution of the Einstein equations with a phantom scalar. The
extra matter, which vanishes for $n=2$, arises when we wish to consider generalisations for $n\neq 2$. 

\noindent One may note that the $\rho_\phi, \tau_\phi, p_\phi$ satisfy the
relation
\begin{equation}
    \rho_\phi=\tau_\phi=-p_\phi = -\frac{1}{2} {\phi'}^2
\end{equation}
where $\phi'=\frac{d\phi}{d\ell}$ and $\phi$ is a phantom scalar with a negative kinetic energy term in the action.\\
Thus, one may sum up and say that the the total action which leads to the
equations of motion of which the $n\geq 2$ line elements are solutions, is given as
\begin{eqnarray}
    S = S_g + S_{ph} + S_{extra} 
    = \int \sqrt{-g}\,  R \, d^4 x  +\int \sqrt{-g} \,\partial^\nu \phi \partial_\nu \phi \, d^4 x \, + \, S_{extra} 
\end{eqnarray}
 where $S_g$ is the standard Einstein-Hilbert gravity action and $S_{ph}$ represents the
 phantom scalar action (note the plus sign (our signature is -+++) 
 which yields the negative kinetic energy for the
 phantom).
  The extra term in the action, $S_{extra}$, is responsible for giving rise to the $n>2$ wormhole geometries and it vanishes for the $n=2$ Ellis-Bronnikov wormhole. Its presence is required because the scalar field alone cannot generate the entire set of $n>2$ geometries. Though we do not know the exact form of $S_{extra}$ (i.e. in terms of a Lagrangian density), we do know, from the field equations, the energy-momentum tensor (the $\rho_e,\tau_e, p_e$
  mentioned above) resulting from  $S_{extra}$ (see eq.(10), (11) and (12)). We will show below how this extra piece of the energy momentum tensor indeed satisfies the ANEC. 
 
 \noindent The equation of motion for the phantom scalar $\phi$ is just $\Box \phi =0$.
 We can solve for the scalar field for arbitrary $n$ metrics to get 
\begin{equation}
\phi(x)= {\sqrt{2}}\, x \,\, {}_2F_{1} \left [1/n, 2/n, 1 + 1/n, -x^n \right ]
\end{equation}

\begin{figure}[h]
 \centering
	\includegraphics[scale=0.85]{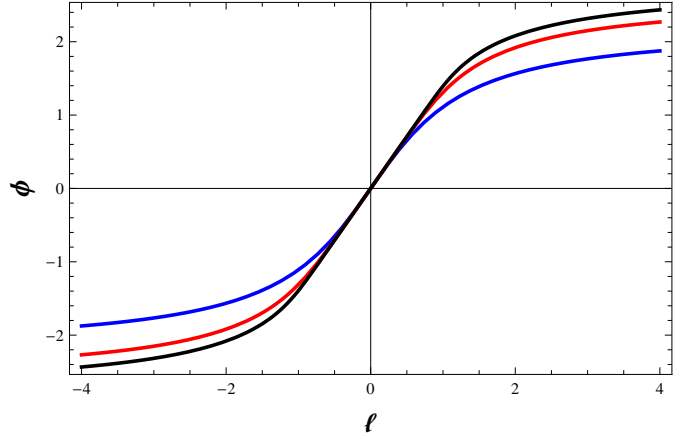}
 	\caption{$\phi$ for $n=2$ (blue), $n=4$ (red), $n=10$ (black)}
 	\label{sclr}
 \end{figure}
\noindent If $n=2$ one obtains $\phi(x) = {\sqrt{2}} \arctan \,x$, which is the solution for Ellis--Bronnikov spacetime. The general solution with
the hypergeometric function has a behaviour similar to the $\arctan \,x$ solution for
$n=2$. This feature can be noted in the graphs in Figure 1, for different $n$ and $b_0=1$.
\noindent It is easily seen that the Euclidean
trace $\rho+\tau+2p=0$ (follows from $R_{00}=0$, stated earlier). If we check 
the WEC inequalities: $\rho\geq 0$, 
$\rho+\tau \geq 0$ and $\rho+p\geq 0$, for the $\rho$, $\tau$ and $p$ stated above, it can be seen that the second inequality is always violated for any value of parameters $n$ and $b_{0}$ as well as $\ell$. The other inequalities may not be violated for certain values of the parameters and over restricted domains of the $\ell$ coordinate. This can be observed more clearly when we plot $\rho (\ell)$, $\rho (\ell) +\tau (\ell)$ and $\rho(\ell) + p(\ell)$ with respect to $\ell$.
 
\begin{figure}[h]
\begin{subfigure}{0.4\textwidth}
	\includegraphics[width=0.95\linewidth]{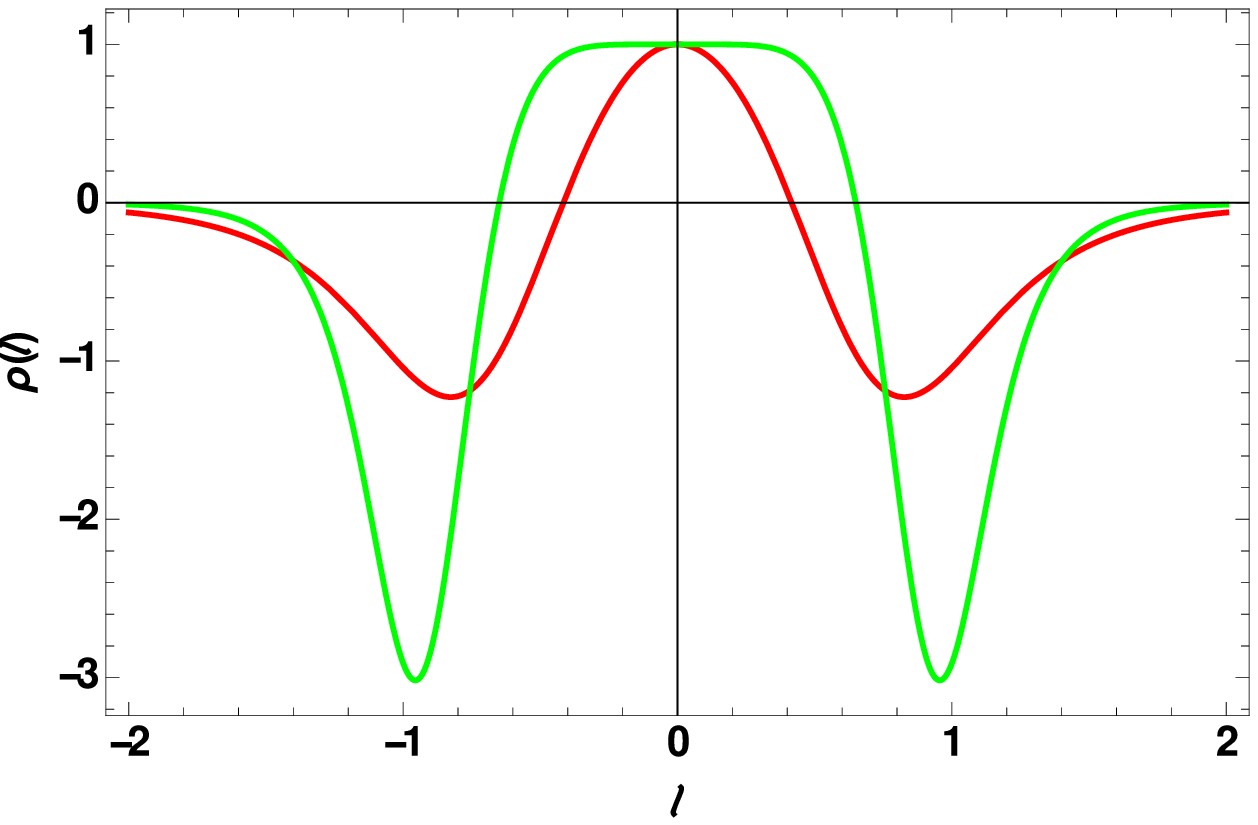}
 	\caption{$\rho (\ell)$ with $\ell$}
 	\label{fig:EC_1}
 \end{subfigure}
 \begin{subfigure}{0.4\textwidth}
 	\includegraphics[width=0.95\linewidth]{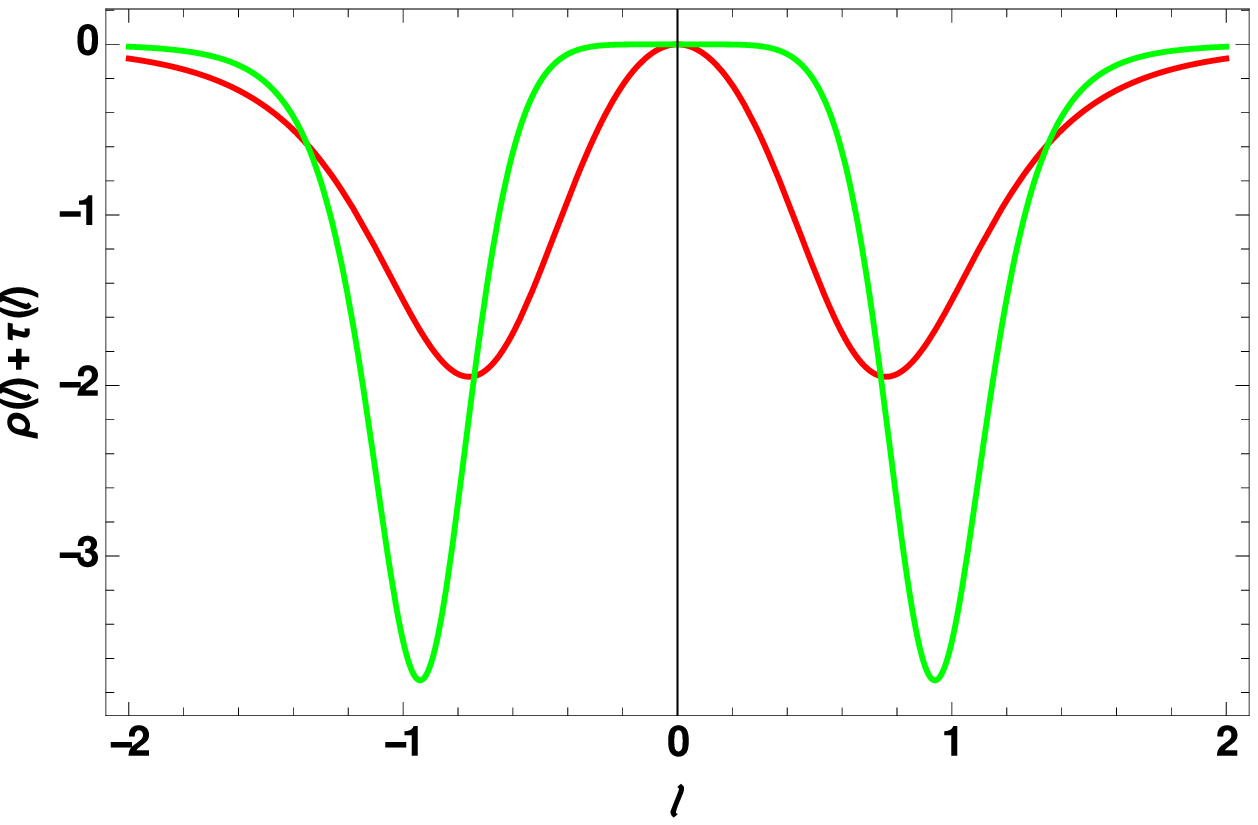}
 	\caption{ $\rho (\ell) +\tau (\ell)$ with $\ell$}
 	\label{fig:EC_2}
 	\end{subfigure}
 \begin{subfigure}{0.4\textwidth}
 	\includegraphics[width=0.95\linewidth]{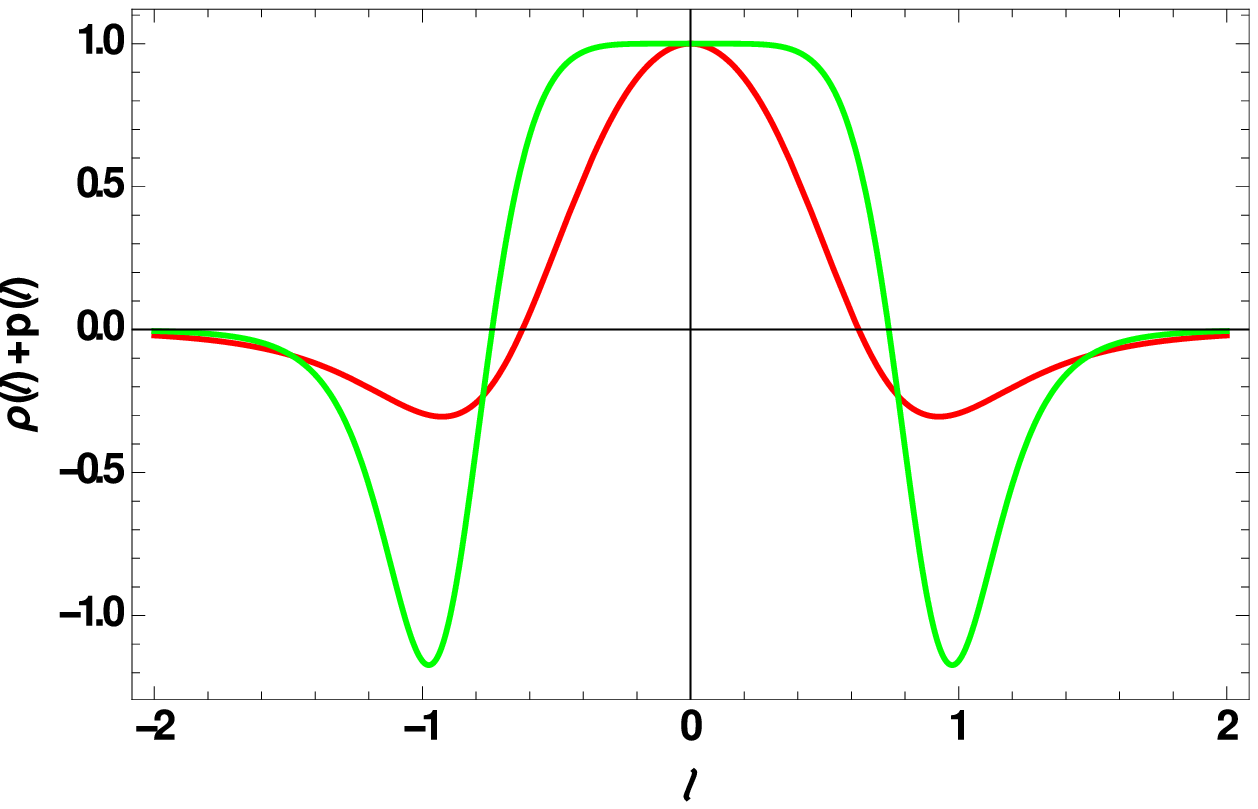}
 	\caption{ $\rho(\ell) + p(\ell)$ with $\ell$}
 	\label{fig:EC_3}
 	\end{subfigure}	
 	\caption{Variation of $\rho (\ell)$, $\rho (\ell) +\tau (\ell)$ and $\rho(\ell) + p(\ell)$ with respect to $\ell$. The red line is for $n=4$ while the green one is for $n=8$ geometry.}
 \end{figure}
\noindent We find that for a range of $\ell$ we do get matter with positive energy density, as can be seen from Fig.(\ref{fig:EC_1}). The third energy condition is also partially satisfied (see Fig.(\ref{fig:EC_3})) but as already stated, the second energy condition is always violated (Fig.(\ref{fig:EC_2})) except at $\ell =0$ (for $n>2$). Since the
$\rho (\ell) +\tau (\ell)$ and $\rho(\ell)+p(\ell)$ inequalities are violated the Null Energy Condition (NEC) is violated as well.
 
 \noindent It is clear from the expressions for $\rho_\phi$, $\tau_\phi$ and $p_\phi$
 that $\rho_\phi \leq 0$, $\rho_\phi +\tau_\phi \leq 0$ and $\rho_\phi + p_\phi=0$, where
 the equality in the first two inequalities happens only at the asymptotic infinities. Thus,
 the phantom scalar field violates the energy conditions everywhere. On the other hand,
 the additional matter given via $\rho_e$,$\tau_e$ and $p_e$ has, as we shall see below,
 the property that it can satisfy the Averaged Null Energy Condition (ANEC) along
 radial null geodesics. 
 
 \noindent The ANEC integral, as is well--known is evaluated along null curves
 with tangent vector $k^i$, and is given by:
 \begin{equation}
     I= \int_{\lambda_1}^{\lambda_2} T_{ij}k^i k^j d\lambda
 \end{equation}
The ANEC is therefore stated as $I\geq 0$.

\begin{figure}[h]
 \centering
	\includegraphics[scale=0.95]{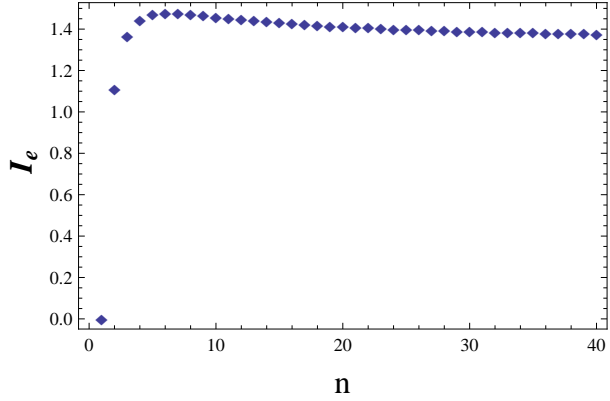}
 	\caption{$I_e$ vs. $n$ }
 	\label{fig:anec}
 \end{figure}

\noindent For our line element, it is easy to see that $(\dot t, \dot \ell, \dot \theta, \dot \phi) = (1,1,0,0)$ represents  null geodesics. If we evaluate the ANEC integral along this
set of null geodesics, we need to work out the integral (choosing $\ell=\lambda$ as the
parameter labeling points on the null geodesics):
\begin{equation}
    I_e = \int_{-\infty}^{\infty} \left (\rho_e +\tau_e\right )\, d\ell
\end{equation}
Using the expressions for $\rho_e$ and $\tau_e$ given in (10) and (11), we find that (for even $n\geq 2$),
\begin{equation}
    I_e = \frac{4 (1-n)}{n^2} \pi \csc \left [\frac{\pi}{n} \right ] + 
  4\frac{\Gamma[1 + \frac{1}{n}] \, \Gamma[\frac{3}{n}]}{\Gamma[\frac{4}{n}]}
\end{equation}
This expression is manifestly positive (see Fig.(\ref{fig:anec})) for all even $n$  and has an asymptotic value equal to $\frac{4}{3}$.
As is well-known, a phantom scalar is a source for the $n=2$ Ellis-Bronnikov geometry. 
However, for $n>2$, the phantom
scalar alone cannot generate the geometries. Additional matter is required, which, interestingly
satisfies the ANEC. This is actually the reason behind the fact that the violation at the throat
does not occur for the $n> 2$ geometries. The fact that the `flaring out' does not happen at the throat
is related to this additional ANEC satisfying matter.
This aspect which is related to the shape of the geometry will become 
clearer through the embedding discussed below.

\subsection{\bf Embedding}
\noindent In order to visualise the shape of such generalised wormholes in 3+1 dimensions, for different values of $n$, the 2-D spatial slice ($t=constant$, $\theta=\pi/2$) of the spacetime is embedded in 3-D Euclidean space with the line element written in cylindrical coordinates. 
To begin, we consider the $t=constant$, $\theta=\pi/2$ slice given as,
\begin{align}
    ds^{2} = d\ell^{2} + (\ell^{n}+b_{0}^{n})^{2/n} d\phi^{2}.
    \label{eq:2D_slice}
\end{align}
The metric has now been reduced to a 2-D geometry which can be visualised by embedding it in a 3-D flat space. Since there is axial symmetry we use cylindrical coordinates ($\zeta, \psi, z$) with the metric of the flat space being
\begin{align}
    ds^{2} = d\zeta^{2} + \zeta^{2} d\psi^{2} + dz^{2}.
\end{align}
As the surface possesses axial symmetry, $\psi=\phi$, $z=z(\ell)$ and $\zeta=\zeta(\ell)$. Thus we get,
\begin{align}
    ds^{2} = \Big[ \Big(\frac{d\zeta}{d\ell}\Big)^{2}+\Big( \frac{dz}{d\ell}\Big)^{2}\Big] d\ell^{2} + \zeta^{2} d\phi^{2}.
\end{align}

\noindent Comparing this with eq.(\ref{eq:2D_slice}) we find,
\begin{gather}
    \zeta (\ell) = (\ell^{n} + b_{0}^{n})^{1/n}
    \label{eq:rho_l}\\
    \frac{dz}{d\ell} = \Big( 1- (\ell^{n} + b_{0}^{n})^{\frac{2}{n}-2} \ell^{2n-2} \Big)^{1/2}.
    \label{eq:z_l}
\end{gather}
The equation (\ref{eq:z_l}) is numerically integrated with $b_{0}=1$ and $z$ is plotted with respect to $\zeta$ in a parametric plot giving the embedding diagram for different values of parameter $n$.

 \begin{figure}[ht]
      \centering
      \includegraphics[scale=1]{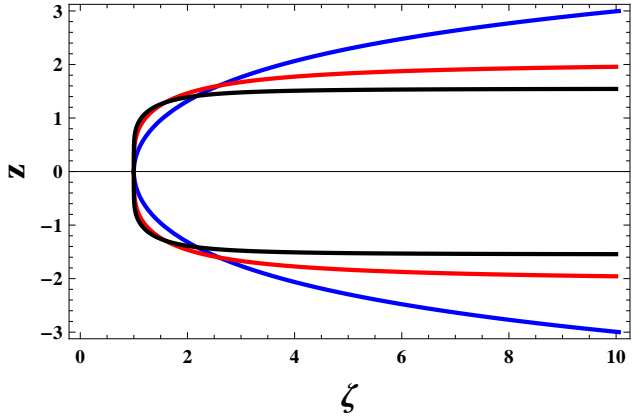}
      \centering
      \caption{Embedding diagram for different values of $n$ ($n=2$(blue), $n=4$ (red), $n=6$ (black)).}
      \label{fig:embed}
  \end{figure} 
  
\noindent As is evident in the plot of Fig.(\ref{fig:embed}), the geometries for different $n$ are quite distinct from each other. It is also observed that with increasing value of $n$ the flaring out of the wormhole right from the throat becomes less prominent. For larger $n$ values 
the embedding will resemble a uniform tunnel connecting two remote flat regions. 
One may also name these geometries as {\em long necked wormholes}.

\begin{figure}[h]
      \centering
      \includegraphics[scale=0.95
      ]{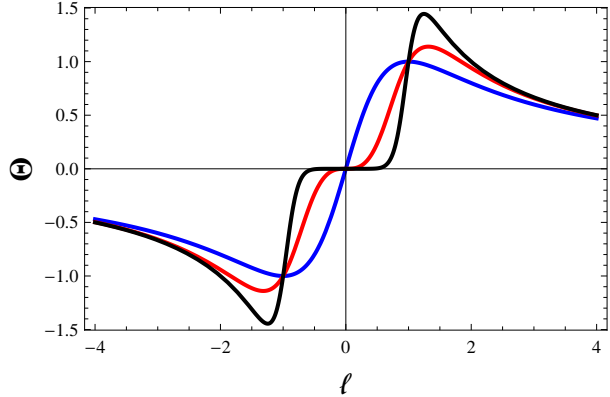}
      \centering
      \caption{Expansion $\Theta$ for different $n$ (n=2 (blue),4 (red),10 (black)).}
      \label{fig:expansion}
  \end{figure}

\noindent The change in shape of the geometry with increasing $n$ is also manifest in the
behaviour of the expansion of a null geodesic congruence. It can be shown that the
expansion $\Theta$ of a null geodesic congruence ($k^i = (1,1,0,0)$) is equal to $2 \frac{r'}{r}$. Evaluating $\Theta$ for $n=2$
one notices that it is zero at the throat but changes significantly 
for $\ell>0$ and $\ell<0$. In contrast, for $n>2$, as shown in Fig.(\ref{fig:expansion}), there is a domain around $\ell=0$
where the expansion is almost zero (exactly so only at $\ell=0$). Thus, for large $n$, a converging
null geodesic congruence entering from one universe, becomes almost parallel ($\Theta\sim 0$ ) over this 
extent of $\ell$, near the throat, before it
starts diverging to become zero again in the `other universe'.  

\noindent In the subsequent sections we will try to see, to what extent, the
nature of the geometries for different $n$ is reflected in the quasinormal modes and
in echoes. 

\section{\bf Scalar wave propagation and effective potentials} 

\noindent Any open system under perturbation emits radiation and looses its energy through a discrete set of complex frequencies of the form $\omega_{QNM} = \omega_{r} + i \omega_{i}$ called the quasi-normal modes \cite{bhm_WH_2,konoplya,kim}. The QNMs are associated with specific boundary conditions, i.e. purely outgoing waves at spatial infinities. The real part of $\omega_{QNM}$ determines the oscillation frequency while the imaginary part gives the damping rate of the field over time. The QNMs can be excited due to scalar, vector (electromagnetic) or tensor perturbations of a given geometry. More details
on the mathematical definition of QNMs and how to find them appears later in this section.

\noindent Let us now consider the propagation of a massless, minimally coupled scalar field in our wormhole spacetimes. This scalar field may be viewed as a perturbation of the phantom scalar which was part of the source for the metric. In other words, we 
may write $\phi +\Phi$ as the perturbed phantom scalar and note that $\Phi$ will
satisfy the same equation $\Box \Phi=0$ as for $\phi$ itself. We intend to calculate the corresponding scalar quasinormal modes (QNMs).
The scalar QNMs will help us in understanding the stability of the spacetime under scalar perturbations and also give us an idea about whether we can distinguish the different geometries of the wormholes through the values of the scalar QNMs. \\
The Klein-Gordon equation for the scalar field $\Phi$ is,
 \begin{align}
    \Box \Phi = 0.
    \label{eq:KG}
\end{align}
As our background spacetime is spherically symmetric and static we use the following ansatz to decompose $\Phi$ in terms of spherical harmonics, where the indices of $Y(\theta, \phi)$ have been suppressed for simplicity.
\begin{align}
  \Phi(t,r,\theta,\phi)=Y(\theta,\phi)\frac{u(r)e^{-i \omega t}}{r}.
\end{align} 
Incorporating this in eqn.(\ref{eq:KG}) we get the radial equation in the form of a Schr\"{o}dinger-like equation in the tortoise coordinate $\ell$,
\begin{align}
  \frac{d^2 u}{d\ell^2}+[\omega^2 - V_{eff}]u=0
  \label{eq:radial_eqn}
 \end{align}
where
\begin{align}
    V_{eff}(\ell)=\frac{m(m+1)}{(b_0^n+\ell^n)^{2/n}}+\frac{(n-1) b_0^n \ell^{n-2}}{(b_0^n+\ell^n)^2}
    \label{eq:potential}
\end{align}
and $m$ is the azimuthal number arising from the separation of variables. For a general $r(\ell)$, it can easily be shown that
\begin{align}
    V_{eff}(\ell)=\frac{m(m+1)}{r^2}+\frac{r''}{r}
    \label{eq:potential_r}
\end{align}
Thus, for $m=0$ (the $s$-wave), $V_{eff} = \frac{r''}{r}$.

\subsection{\bf Effective potentials, energy conditions and single/double barriers}

\noindent An interesting fact to note about the $m=0$ effective potential
(for any $r(\ell$) stated just above) is that it is linked to
$\rho(\ell)$ and $\tau(\ell)$ via the Einstein equations for a general $r(\ell)$.
We can easily note,
\begin{align}
    V_{eff}(\ell) =\frac{r''}{r} =-\frac{1}{2} \left (\rho+\tau\right ) = p
\end{align}
Thus, to have an everywhere non-negative effective potential, we need to have 
$\rho+\tau\leq 0$ or $p\geq 0$.
The single or double barrier nature of the effective potential depends on the
number of zeros of $\frac{dp}{dr}$ and their nature (maximum/minimum etc.). 
If $\frac{dp}{dr}$ has only one zero which is a maximum and since $V_{eff}$ goes to zero asymptotically from the positive side if $\rho+\tau \leq 0$, we have a single barrier. 
On the other hand, if $\frac{dp}{dr}$ has three zeros (two maxima and one minimum) then
we can have a double barrier, assuming $r(\ell)$ to behave as required for a wormhole.
Thus, choosing a $r(\ell)$ and ensuring that $\frac{dp}{dr}$ has the right behaviour
we can generate wormholes which will have a double barrier effective potential.

\noindent It is important to note that our $n>2$ geometries necessarily have these
properties. One can verify that the extrema for any value of $n$ occurs at
\begin{align}
    \ell = 0, \pm \left (\frac{n-2}{n+2}\right )^{\frac{1}{n}} b_0
\end{align}
where $\ell=0$ is a minimum with $r''(0)=0$ and the other two symmetrically placed (about $\ell=0$)
locations are maxima.
Many more examples can be worked out which exhibit double barriers. For example
$r(\ell) = \sqrt{\ell^2 + b_0^2 e^{-\alpha \ell^2}}$ will yield a double barrier too for $\alpha \neq 0$
and it will reduce to Ellis--Bronnikov for $\alpha=0$. It can be checked that none of the three standard
wormhole spacetimes: spatial Schwarzschild ($b(r)=b_0$), Ellis--Bronnikov ($b(r)= \frac{b_0^2}{r}$)
and the wormhole with $b(r) = \left (b_0\right )^\frac{1}{n} r^{1-\frac{1}{n}}$ ($n>1$) exhibit
double barriers in their scalar wave effective potentials.

\noindent Thus, energy condition violation and the existence of multiple zeros and
extrema of the gradient of the tangential pressure are the decisive factors in 
generating an everywhere positive, double barrier, $m=0$ effective potential. The existence of
the double barrier is crucial for the phenomenon of echoes to occur, as we discuss later.

\noindent One may further ask, what happens when $m\neq 0$? In that case the
$V_{eff}$ has an additional term $\frac{m(m+1)}{r^2}$, which is manifestly positive.
Does it change the nature of the potential? One can verify that the maxima locations
$x_{max}$, now emerge from the roots of the relation ($n$ even):
\begin{equation}
    \left (x_0^n-x_{max}^n \right )^{\frac{n}{2}} = \alpha^{\frac{n}{2}} x_{max}^{n} \left (1+x_{max}^n
    \right )^{n-1} 
\end{equation}
where $x_{max}=\frac{{\ell}_{max}}{b_0}$, $x_0 =  \left (\frac{n-2}{n+2}\right )^{\frac{1}{n}}$ 
(maxima for $m=0$)
and $\alpha = \frac{2m(m+1)}{(n-1)(n+2)}$. It is easy to note from the above expression 
that $x_{max}^2 < x_0^2$ or $\vert x_{max}\vert < \vert x_0\vert$. This means the maxima 
shift towards the origin as $m$ becomes larger and the value of $V_{eff}$ at $\ell=0$ is 
gradually lifted upwards. For large $m$, the double barrier features almost
(not fully) disappear and we end up being closer to a single barrier located at $\ell=0$.
However, if $n=2$, $x_0=0$ and the above equation has no solution apart from 
$x_{max}=0$, which is a maximum for all $m$.

\begin{figure}[h]
\begin{minipage}[t]{0.45\textwidth}
      \includegraphics[width=\textwidth]{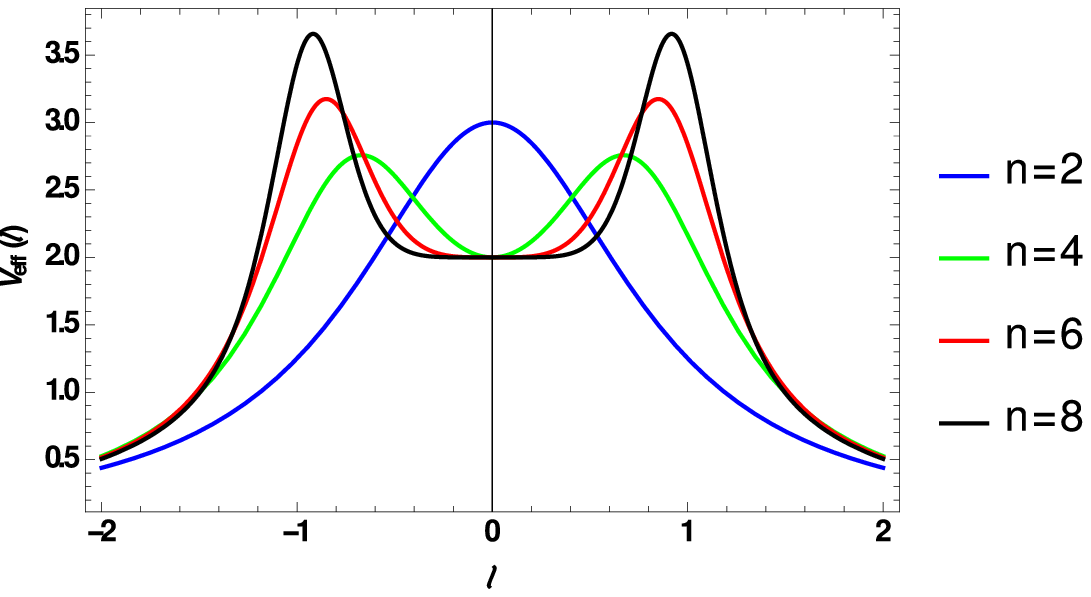}
      \caption{Potentials for different values of $n$ for $m=1$, $b_0=1$.}
      \label{fig:potential}
      \end{minipage}
      \hspace{0.25in}
\begin{minipage}[t]{0.45\textwidth}
 \includegraphics[width=\textwidth]{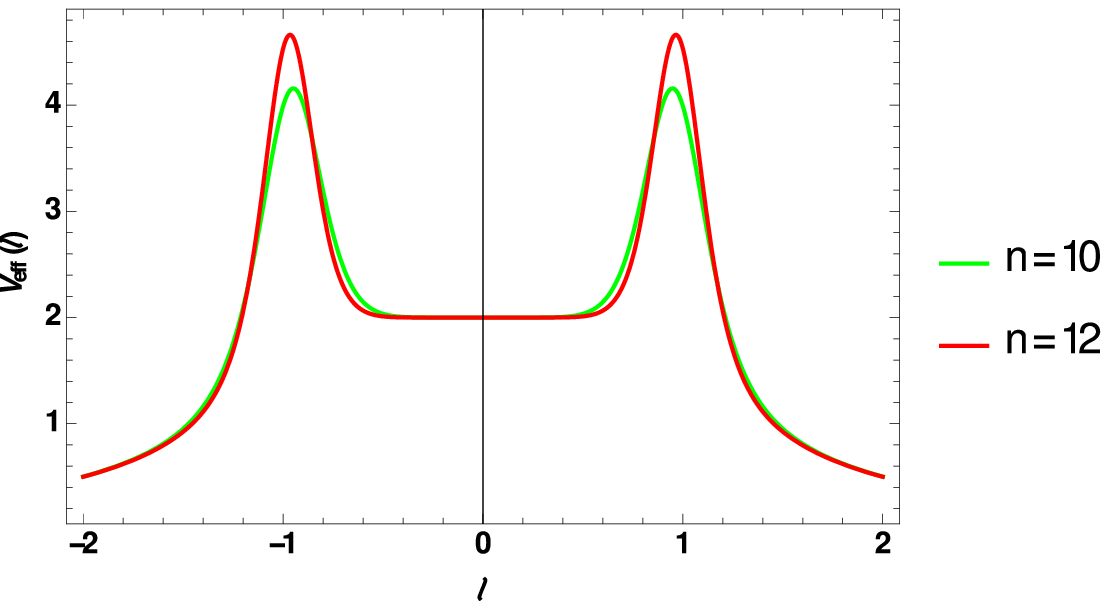}
  \caption{Potentials for higher values of $n$ and $m=1$,$b_0=1$.}
  \label{fig:higherpotential}
\end{minipage}
\end{figure}

\noindent We state below some features of the effective potential with reference to the associated figures.

\noindent $\bullet$ The plot in Fig.(\ref{fig:potential}) shows the variation of $V_{eff}(\ell)$ w.r.t $\ell$ for different geometries at a particular value of $m = 1$ and $b_0=1$. It can be seen that the potential has a single barrier feature only for $n=2$, for all $n>2$ there are double barriers. This is true for {\em lower values of $m$}. Hence the $n>2$ geometries are completely different from the Ellis-Bronnikov geometry.

\noindent $\bullet$ The plot of the potential (Fig.(\ref{fig:higherpotential})) for {\em higher $n$ geometries}, even for lower modes, shows that although the potential is still a double barrier, the potential curves for different geometries are nearly identical. 

\begin{figure}[h]
\begin{minipage}[t]{0.45\textwidth}
 \includegraphics[width=\textwidth]{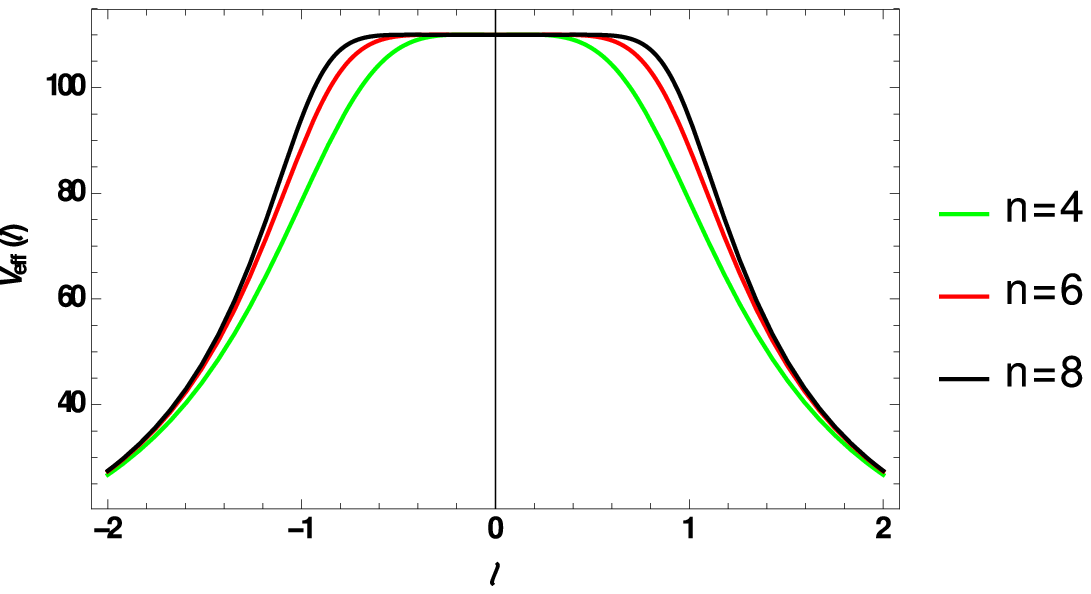}
  \caption{Potentials for different values of $n$ for $m=10$, $b_0=1$.}
  \label{fig:potential1}
\end{minipage}
\hspace{0.3in}
\begin{minipage}[t]{0.45\textwidth}
 \includegraphics[width=0.85\textwidth]{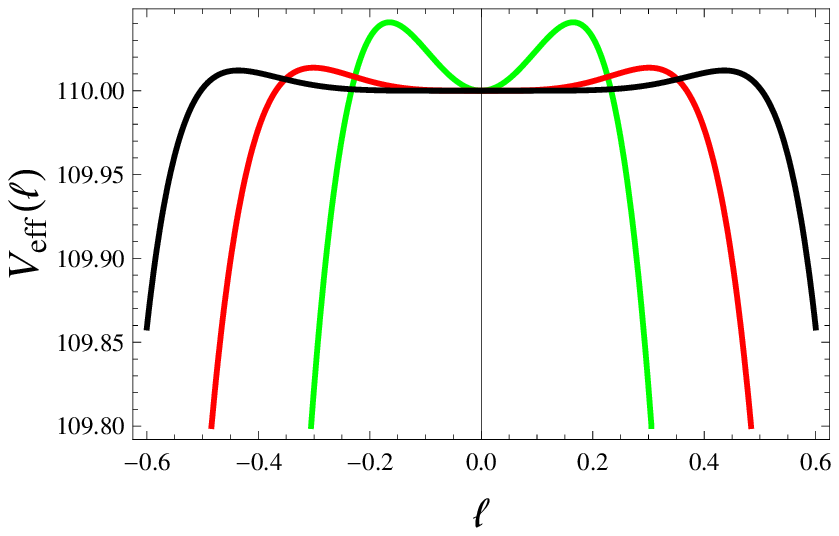}
  \caption{`Nearly' single barrier potential for $n=4$(green), $n=6$(red), $n=8$(black); $m=10$,  $b_0=1$.}
  \label{fig:near_singlepotential}
\end{minipage}
\end{figure}

\noindent $\bullet$ In Fig.(\ref{fig:potential1}), we find the potential being plotted once again for different geometries but with {\em higher $m$ value}. It is interesting to observe that here the potential, for all geometries (i.e. for all $n>2$), show an almost single barrier structure similar to the $n=2$ case. We can see the two peaks getting flatter and merge into a `nearly' single barrier when we zoom in very close to $\ell=0$ (see Fig.\ref{fig:near_singlepotential} for a zoomed plot of the effective potential--note the y-axis range and the scale here).

\noindent $\bullet$ For all the potential plots we have taken the throat radius as unity. From eq.(\ref{eq:potential}) if we write the potential in terms of $x = \ell/b_0$, we observe that the $V_{eff}$ varies as $b_0^{-2}$. So for higher values of $b_0$ the height of the potential peak will go on decreasing. Also, for any $n>2$ geometry the width of the peak as well as their separation (for small $m$) increases for higher $b_0$ values.

\vspace{0.1in}

\section{Quasinormal modes}

\begin{figure}[h]
 \centering
\begin{subfigure}[t]{0.47\textwidth}
  \centering
	\includegraphics[width=\textwidth]{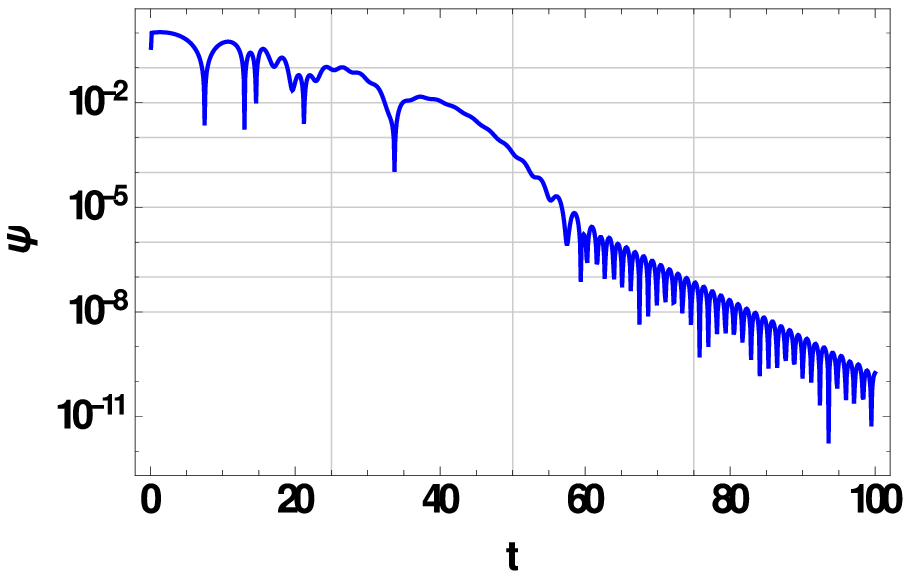}
 	\caption{ $n=4,m=2,\ell=7$}
 	\label{fig:TD_1}  
 \end{subfigure}
 \begin{subfigure}[t]{0.47\textwidth}
 	\centering
 	\includegraphics[width=\textwidth]{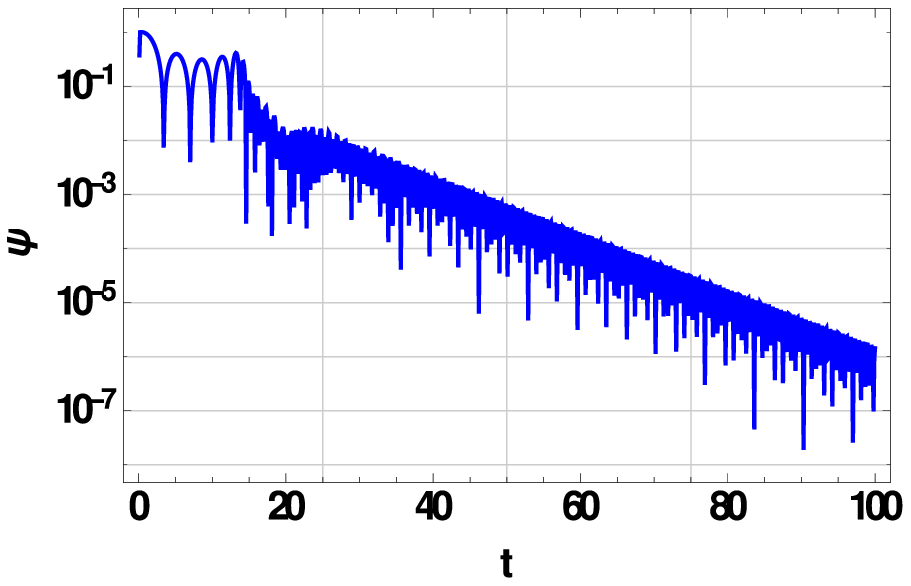}
 	\caption{$n=6,m=5,\ell=7$}
 	\label{fig:TD_2}
 	\end{subfigure}
 	\caption{Time domain profiles showing the characteristic QNM ringing in semi-logarithmic scale. The integration is done on a $u-v$ null grid with a step size 0.1. The initial conditions are defined on the lines $v=0$ as a Gaussian profile centered at $u=10$, $\psi(u,0)= e^{\frac{-(u-10)^{2}}{100}}$ and, on the $u=0$ line, as a constant, $\psi(0,v)= constant$, the value of which is determined by $\psi(0,0)$. For details on the discretization scheme, see \cite{konoplya_TD}.}
 \end{figure}

\noindent Given the features of the effective potentials,  we now move on towards obtaining the scalar quasinormal modes in the
background geometry of this family of wormholes. 
The existence of QNMs can be directly observed in the time evolution of the scalar field obtained by integrating the scalar wave equation following methods outlined in \cite{price, konoplya_TD}. The wave equation
\begin{align}
    \frac{\partial^{2}\psi}{\partial t^{2}} - \frac{\partial^{2}\psi}{\partial \ell^{2}} + V_{eff}(\ell) \psi =0 .
    \label{eq:QNM}
\end{align} 
is recast using light cone coordinates ($du = dt- d\ell$ and $dv = dt + d\ell$). 
Using appropriate initial conditions along the $u$ and $v$ lines we numerically integrate
to obtain the time-domain profiles shown in Fig.(\ref{fig:TD_1}) and Fig.(\ref{fig:TD_2}). The damped ringing in time, exhibiting the decay of the scalar field is clearly visible in
these plots. 

\noindent In order to find the quasinormal modes we use various available methods.
Let us first discuss the methods briefly. Thereafter, we
present our results and discuss their consequences in detail.

\noindent 
\subsection{Methods for finding QNMs}

\noindent As stated before, QNMs are complex frequencies associated 
with purely outgoing waves at spatial infinity. From
eq.(\ref{eq:QNM}) we see that $\psi(\ell,t) \rightarrow e^{\mp i \omega \ell} e^{-i \omega t}$ as $\ell \rightarrow \mp \infty$  (note,
for an asymptotically flat spacetime, $
V_{eff} \rightarrow 0$ at spatial infinity).
Thus, if a QNM frequency has a negative imaginary part, the field will decay with time through these modes indicating a stable geometry.

\noindent  
We may find
the scalar QNMs for different geometries using three different methods. The method of Prony fitting and direct integration are completely numerical techniques while the WKB method is semi-analytic. In the following, we discuss the direct integration method in detail and refer to well-known sources for the other two methods.\\

\noindent The time domain evolution of a signal  can be used to extract the QNMs by fitting damped exponentials through the method of Prony fitting. 
This method has been applied multiple times in the
literature (for details see \cite
{konoplya_TD}).  We get a precision of upto 3 decimal places through this method.
A possible source of error here is the lack of knowledge about 
the exact beginning of QNM ringing. However,
this error does not really affect our results as we are interested only in the dominant mode obtained by fitting the late-time part of the signal with damped exponentials.\\
\noindent Another method for finding the QNMs is the semi-analytical method or WKB approximation which was developed by Schutz and Will \cite{schutz}. They calculated the QNM frequencies by taking the WKB solutions upto the eikonal limit which gives a simple analytical formula  involving the parameters in the metric functions,
\begin{align}
     \omega^{2} = V_{0} - i (p+\frac{1}{2}) \sqrt{-2 V''_{0}}, \hspace{0.2in} p=0,1,2....
\end{align}
where $V_{0}$ and $V''_{0}$ denote the values of the effective potential and its second derivative at the maximum. $p$ denotes the overtone number with $p=0$ being the fundamental mode. In our work we have dealt with only the fundamental modes and compared 
some of the WKB values with the numerically obtained results. 
For a recent comprehensive review on WKB methods one may refer to \cite{zinhalio}. 
We note that the WKB method is not perturbative. Hence, higher orders may not necessarily ensure better results.  Also, one needs to keep in mind that the WKB formula used here is applicable for single barrier potentials i.e. when two turning points are involved. In our case though, we have double barriers which tend to almost single barriers for large $m$. Thus, WKB results will not be applicable in general (may be so, in an approximate sense, for large $m$) and the formula needs to be suitably modified for four turning points in order to obtain correct results. 
\noindent Finally, we discuss the direct integration method for finding QNMs, first given by Chandrasekhar and Detweiler \cite{Chandrasekhar}. Here, the differential equation (\ref{eq:radial_eqn}) is numerically integrated using purely outgoing boundary conditions. Since the potential is symmetric about the throat of the wormhole at $\ell=0$, the solutions will be symmetric or anti-symmetric. Thus, we can impose additional conditions on the solutions namely: $u(0)=0$ for anti-symmetric solutions and $u'(0)=0$ for symmetric solutions. This gives us two classes of QNMs corresponding to the condition imposed at $\ell=0$.\\
We observe that only $e^{-i \omega \ell}$ wave will exist on the left side of the throat as there is no reflecting potential at $\ell = - \infty$. Hence the throat of the wormhole seems to play the role of the event horizon \cite{kim}. Ideally, we should integrate from $\ell = -\infty$ to $\infty$. However, since our potential is symmetric about $\ell=0$
we integrate in the positive half range of $\ell$ corresponding to $r=b_0$ ($\ell=0$) to $\infty$ and then reflect the solution about the throat by using the conditions
for obtaining symmetric or anti-symmetric solutions. In this way, we are taking into account the behavior of the solution for $\ell<0$ and we have the right boundary
conditions (outgoing) at both the infinities in $\ell$ (i.e. $\ell\rightarrow \pm\infty$). 
A similar treatment can be found for another wormhole geometry in \cite{aneesh}.

\noindent We begin with the series expansion of $u$ at infinity
given as,
\begin{align}
    u = e^{i\omega \ell} \sum_{n=0}^{N} \frac{A_{n}}{\ell^{n}}; \hspace{0.2in} \ell \rightarrow \infty.
\end{align}
and obtain the coefficients $A_{1},A_{2},...$ in terms of $A_{0}$. We integrate the differential equation (\ref{eq:radial_eqn}) from $\ell=0$ to an arbitrary point $\ell_0$ (which lies between the throat and infinity) with the initial conditions at the throat as $u(0)=0$ for anti-symmetric and $u'(0)=0$ for symmetric solutions. 
At $\ell_{0}$, the solution will have contribution from $e^{\pm i \omega \ell}$ terms i.e. both outgoing and ingoing waves will contribute. 
\begin{align}
    u = e^{i\omega \ell} \sum_n \frac{A_{n}}{\ell^{n}} + e^{-i\omega \ell} \sum_n \frac{A'_{n}}{\ell^{n}}
    \label{eq:expansion}
\end{align}
Now, $u$ can be written in terms of only $A_{0}$ and $A'_{0}$. The integrated solution and the expansion (\ref{eq:expansion}) must match along with their derivatives at the point $\ell_{0}$. If we set the point $\ell_{0}$ to $\infty$, then, as per the boundary condition of QNMs, only purely outgoing solutions will survive at infinity i.e. $A'_{0}=0$, the roots of which will give us the values of the QNM frequencies. The stability of the method can be checked by changing the matching point which should not affect the QNM value.
In our work, we will consider QNM frequencies related to the symmetric class of
solutions, which have low damping. 

\noindent It may be noted that the error in this method arises due to the fact that the boundary condition imposed is not exactly at spatial infinity but at some finite point. 
This error may be minimised if we choose the matching point to be sufficiently far 
away from the throat. 
In our quoted values, we have a precision of six(6) digits for the imaginary part of the QNMs 
and five(5) digits for the real part.
\subsection{\bf Results from various methods
}
\noindent We have found the QNMs using each of the above numerical methods, assuming the 
throat radius $b_{0}=1$. The following two tables (Table \ref{tab:n4} and Table \ref{tab:n10}) list the fundamental QNM values calculated numerically for different modes, for the $n=4$ and $n=10$ geometries. We observe that the two numerical methods give nearly identical values and the results from both methods are suitable for use in further calculation. However, we use the QNM values from the Prony method in the next section, although using values from DI would lead to exactly 
the same observations.

\begin{table}[H]
\begin{minipage}[t]{0.3\linewidth}
\begin{tabular}{|c|c|c|}
     \toprule[0.8pt]
     m & Prony & DI \\ 
    \hline \hline
    1 &  1.66475-i 0.309338 & 1.68973-i 0.284052\\ 
    
    3  & 3.63213-i 0.219813 &  3.62854-i 0.221482\\ 
    
    5 &  5.62378-i 0.191588 & 5.60815-i 0.195234\\
   
    7  & 7.6388-i 0.172743 &  7.59739-i 0.178749\\ 
    
    10 & 10.70612-i 0.151295 &  10.58773-i 0.161901\\
    \bottomrule[0.8pt]
\end{tabular}
\centering
\caption{\label{tab:n4} $\omega_{QNM}$ values for different modes for $n=4$.}
\end{minipage}
\hspace{1.5in}
\begin{minipage}[t]{0.3\linewidth}
\begin{tabular}{|c|c|c|}
     \toprule[0.8pt]
     m  & Prony & DI \\
     \hline \hline
     1 &  1.72679-i 0.189702 & 1.72666-i 0.190301\\
    
     3 &  3.65001-i 0.110534 &  3.64699-i 0.111753 \\ 
    
     5 &  5.6289-i 0.079654 &  5.61434-i 0.082101\\
    
     7 & 7.63526-i 0.0628301 &  7.59540-i 0.066223 \\ 
    
     10 &  10.69510-i 0.0473238 &  10.5779-i 0.052474\\ 
    \bottomrule[0.8pt]
\end{tabular}
\centering
\caption{\label{tab:n10}$\omega_{QNM}$ values for different modes for $n=10$.}
\end{minipage}%
\end{table}

\subsection{\bf Wormhole shapes from QNMs?}

\begin{figure}[ht]
     \centering
      \includegraphics[scale=1.2]{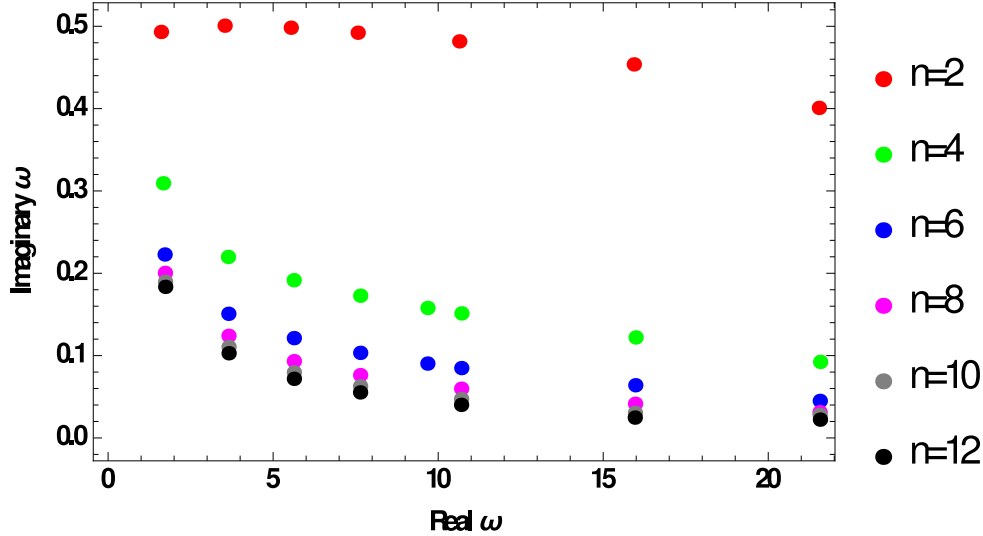}
      \caption{Plot showing variation of real $\omega_{QNM}$ with the magnitude of its imaginary part for different $n$ values. m increases from left to right for each n.}
      \label{fig:QNMPlot}
\end{figure}

\noindent It is now reasonable to ask: is it possible, using QNMs to distinguish between geometries with different $n$ values.
The plot in Fig.(\ref{fig:QNMPlot}) shows the variation of the real part of the fundamental $\omega_{QNM}$ with the magnitude of its imaginary part, for different $n$. Each point in the plot for a particular $n$ corresponds to its QNM frequency for a particular angular momentum mode. As we move from left to right, the value of $m$ goes on increasing, so the left-most point corresponds to lowest $m$, while the rightmost point is for the highest $m$
value. As mentioned before, the QNM values used in this plot have been calculated using the Prony method.
From the plot of QNM frequencies in Fig.(\ref{fig:QNMPlot}) we can draw the following conclusions:\\
\noindent $\bullet$ For lower $n$, the geometries are distinguishable through their fundamental scalar QNMs.\\
$\bullet$ When we move to higher $n$, the geometries begin to look nearly identical as 
evident from the QNMs (and also the effective potentials Fig.(\ref{fig:higherpotential})) for all modes. Hence it becomes difficult to distinguish different higher $n$ geometries, 
solely from the QNMs.\\
$\bullet$ Note that lower $m$ modes are more suited for identifying the geometries. 
Higher $m$ modes of all geometries have similar QNM values due to their nearly identical almost
single barrier effective potentials (see Fig.(\ref{fig:potential1})).\\
$\bullet$ We also observe that as we go to higher $n$ geometries the magnitude of the imaginary part of $\omega$ goes on decreasing. Thus the higher $n$ wormhole families are likely to be relatively less stable owing to the fact that the perturbation for these 
wormholes takes a longer time to decay. \\
$\bullet$ It is also clear that the $n=2$ values are markedly different from those for other $n$. From a geometry standpoint, one is aware that the $n=2$ geometry (Ellis--Bronnikov)
is indeed special and has clear distinguishing features (see earlier discussion in
Section II).

\subsubsection{\bf{Approximate Analytic Fit}}

\noindent In order to extract physically relevant information from our numerically
obtained QNM values, we need an approximate model for $\omega$ which may be obtained by fitting the numerical data to analytical functions.


\noindent We intend to study the variation of frequency (obtained from the real part of $\omega$) with throat radius for different geometries and modes.
We observe from Table \ref{tab:n4} and \ref{tab:n10} that $\omega_r$ increases with increasing $m$. The value of $\omega_r$ also varies inversely with the throat radius. To imitate such a behavior we construct an approximate analytic model,
\begin{align}
    \omega_{r} = c\, \Big(\frac{m\,+\,a\,n^{d}\,+\,g\,n^{2}\,+\,h\,n^{3}\,+\,k\,n^{4}+\,p\,n^{5}}{b_{0}}\Big) = c\, \Big(\frac{f(n,m)}{b_0} \Big)
    \label{eq:fit}
\end{align}
where $c$ is the speed of light, $m$ is the corresponding mode which we want to fit, $b_{0}$ is in length units and the magnitudes of the coefficients $a$, $g$, $h$, $k$, $p$ and the exponent $d$ are obtained using NonLinearModel fit in {\em Mathematica 10}, which fits the above model with frequency corresponding to each mode, for each geometry, as obtained earlier using the Prony method ( Sec.IV). We have used six frequencies corresponding to six values of $n$ for finding the coefficients and the exponent in the fit model. The analytical fits as obtained using $Mathematica $ 10 for $m=1$ and 10 has been discussed in detail in the Appendix. By looking at the explicit values of the coefficients as given in eq.(\ref{eq:fit_realm1}) and (\ref{eq:fit_realm10}) for modes $m=1$ and 10 respectively, we can get an idea about the contribution made by each power of $n$, which is true for all modes. 

 \begin{figure}[h]
  \centering
  \begin{subfigure}[t]{0.45\textwidth}
	\includegraphics[width=1\textwidth]{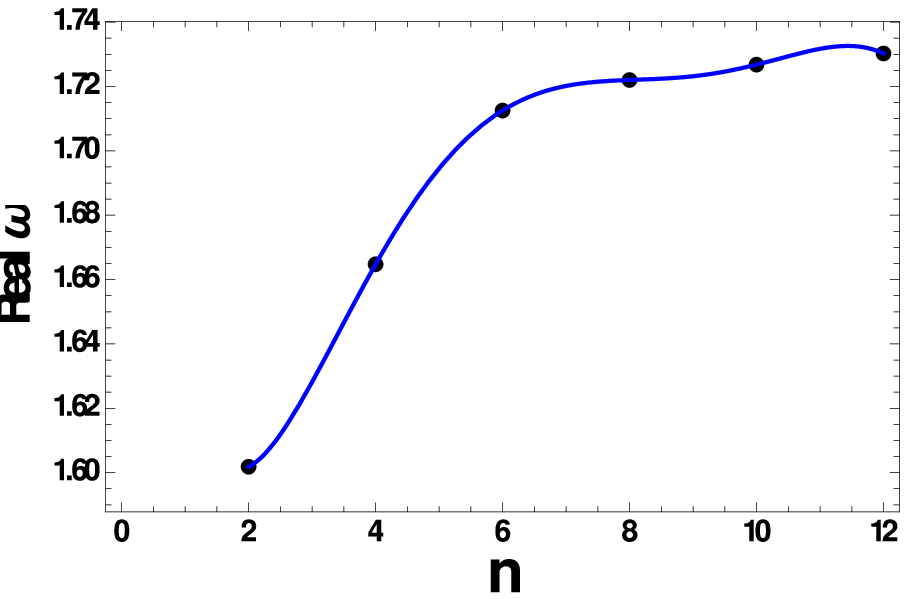}
  	\caption{ For $m=1$}
 	\label{fig:fit_m1}
 \end{subfigure}
  \begin{subfigure}[t]{0.45\textwidth}
 	\includegraphics[width=1.05\textwidth]{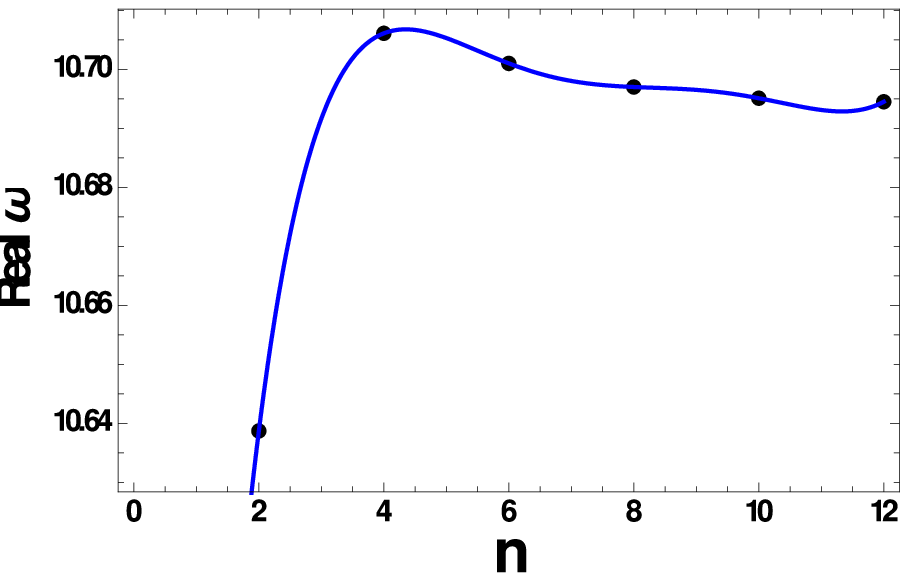}
 	\caption{ For $m=10$}
  	\label{fig:fit_m10}
  	\end{subfigure}
  	\caption{Plot shows how well the real part of $\omega$ matches to the fit relation for different values of $n$. The smooth curve is the analytical fit with $b_0 = 1,\, c=1$.}
  \end{figure}
 
\noindent From Fig.(\ref{fig:fit_m1}) and (\ref{fig:fit_m10}) we note that the best-fit model matches the values of the frequencies quite well. If we refer to the values of the coefficients and the exponent in the fit model (given in the 
Appendix for $m=1$ and 10), we find that, as expected, the higher powers of $n$ contribute much less to the value of the frequency. We can now use the fit model to write the throat radius, in units of $M_{\odot}$ (which is physically meaningful), corresponding to each geometry as
\begin{align}
    \nu = \frac{f(n,m)\, c^3}{2\, \pi \, b_0 \, G \, M_{\odot}} Hz
    \label{eq:freq}
\end{align}
with $f(n,m)$ denoting the fitting function for each mode and geometry, $G=6.67 \times 10^{-11}$ $m^3\,kg^{-1}\, s^{-2}$ and $M_{\odot} = 1.989\times 10^{30} $ kg. 
The accuracy of the fit as given in $Mathematica$ is upto 5 decimal places i.e. the results obtained from the fit exactly matches the ones obtained numerically, upto 5 decimal places. 
The plots shown in Fig.(\ref{fig:fit_m1}) and Fig.(\ref{fig:fit_m10}) have a precision of 5 decimal places. The fitting function (as given in the Appendix) has been used to generate the plots in {\em Mathematica}.\\
Similarly, we can get a model fit for the imaginary part of the QNM frequency as well, to study the damping time. The approximate model can be taken as, 
\begin{align}
    \omega_i = c \Big( \frac{a\, n^d\, + p\, n^2\, +q\, n^3\,+k\,n^4\,+g\,n^5}{b_0\, m}\Big) = \frac{c F(n,m)}{b_0}
    \label{eq:fit_im}
\end{align}
 where the magnitudes of the coefficients $a$, $p$, $q$, $k$ and $g$ and the exponent $d$ are to be determined from the NonLinearModel fit in {\em Mathematica 10}. As before, for the detailed fitting functions for modes $m=1$ and 10, the reader is referred to the Appendix.
The accuracy of this fitting model is the same as stated just above, for the real part. 
The damping time will be given by a relation
\begin{align}
    \tau = \frac{2 \pi}{\mid \omega_i \mid} = \frac{2\,\pi\,b_0\,  G \,M_{\odot}}{c^3\, F(n,m)}.
    \label{eq:damping_time}
\end{align}
The above relation can be used to compute the damping time for any geometry corresponding to any mode for a particular $b_0$.

\begin{figure}[h]
 \centering
\begin{subfigure}[t]{0.45\textwidth}
	\includegraphics[width=1.1\textwidth]{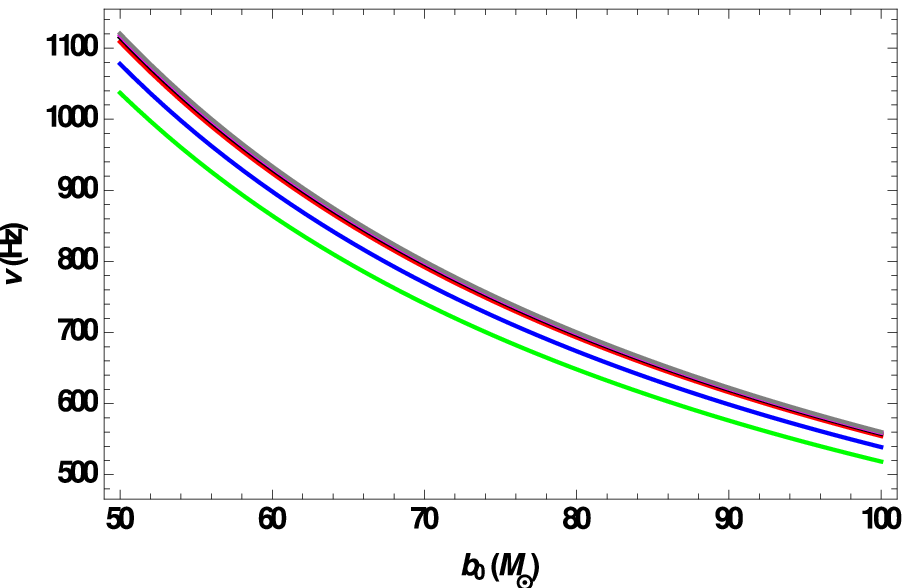}
 	\caption{ For $m=1$}
 	\label{fig:freq_b}
 \end{subfigure}
 \hspace{0.2in}
 \begin{subfigure}[t]{0.45\textwidth}
 	\includegraphics[width=1.3\textwidth]{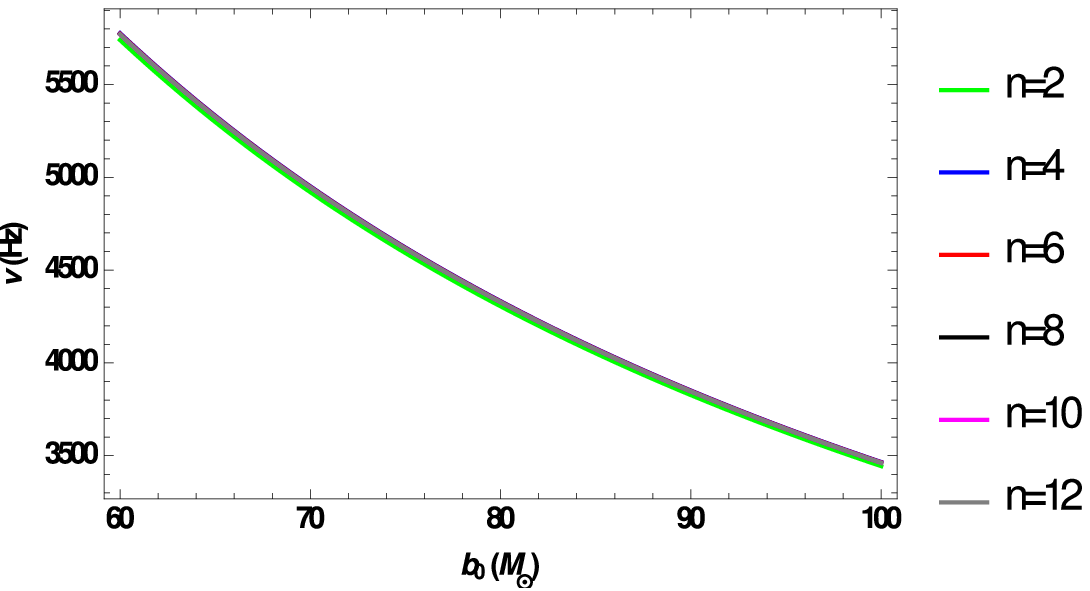}
 	\caption{ For $m=10$}
 	\label{fig:freq_b1}
 	\end{subfigure}
 	\caption{Plot showing the variation of frequency for a range of values of $b_0$ corresponding to different geometries as obtained from the fitting relation in (\ref{eq:freq}).}
 \end{figure}

\noindent We now try to find frequencies, using eq.(\ref{eq:freq}), corresponding to a range of values of $b_0$ while keeping $m$ fixed for each geometry.
The values of the constants as shown in eq.(\ref{eq:fit_realm1}) and eq.(\ref{eq:fit_realm10}) are substituted in $f(n,m)$ (eq.(\ref{eq:freq})) and the frequency is plotted as a function of throat radius. In Figs.(\ref{fig:freq_b}) and (\ref{fig:freq_b1}), we observe that for almost the same range of throat radius, the frequencies corresponding to various geometries have higher values for higher $m$ modes. Hence, it is the lower modes which bear a closeness with the frequencies observable in
the current generation of gravitational wave detectors. In the case of higher $m$ values, the frequencies for geometries with different $n$ are very similar (see Fig.(\ref{fig:freq_b1})) 
over the entire range of the throat radius . Thus, even if we succeed in detecting the higher modes in some way, they will not help us in determining the corresponding geometry of the wormhole from the frequency. In contrast, for the lower $m$ modes, the lower $n$ geometries ($n=2,4,6$) indeed have different frequencies for the same throat radius, thus making them distinguishable.

\noindent It is therefore fair to say that the QNMs can indeed be used as a tool
to identify geometries with different $n$. It seems, for large $n$, in particular,
difficulties appear. However, as we now illustrate, large $n$ has another distinguishing
feature -- the occurence of echoes. 

\section{Observing echoes for large `$n$' values}

\noindent Earlier, we noted that for $n>2$ and small $m$, the effective potential is a double barrier, as shown in Fig.(\ref{fig:potential}). 
In the presence of such a potential, after the initial damped ringdown 
phase (i.e at later times), the transmitted wave emerges after getting 
successively reflected between the two potential peaks. 
Thus, the transmitted waves have a relative time delay along with a
reduction in their peak amplitude. This is the well-known phenomenon
of `echoes in the time-domain profile' which is manifest in
double barrier potentials. These echo signals appear as follow-up to the 
initial ringdown phase and dominate only at late times. They play an important role in distinguishing the spectrum of a black hole from exotic compact objects that are generally characterised by two potential surfaces, thus 
leading to the possibility of echoes. Echoes have been studied for a wide variety of cases, some of which can be found in \cite{kerr_WH,echo1,echo3,echo4,echo5,echo6,echo7,echo8,echo9,echo10,echo11,echo12}.



\begin{figure}[h]
 \centering
\begin{subfigure}[b]{0.45\textwidth}
	\includegraphics[width=1.1\textwidth]{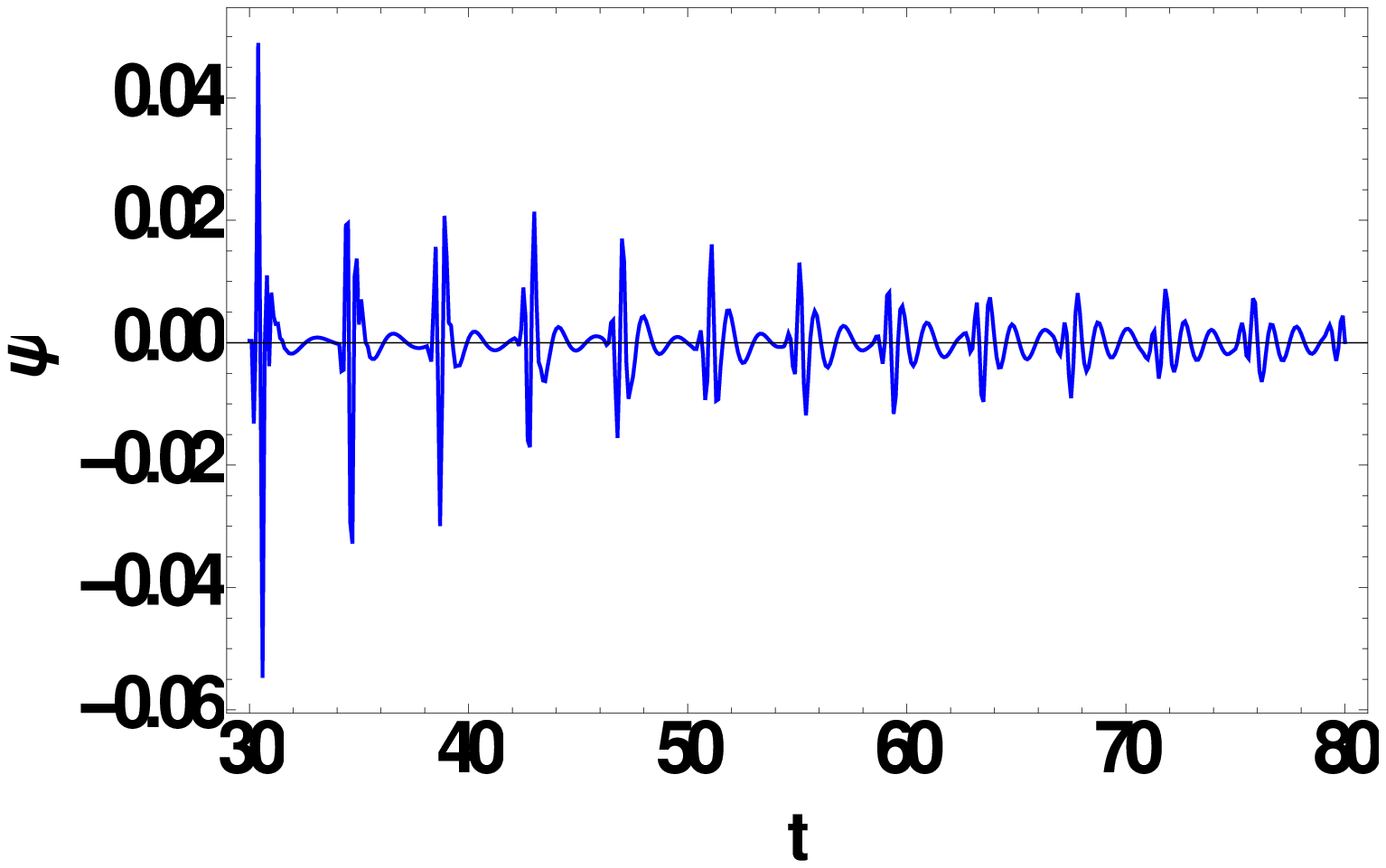}
 	\caption{ For $n=2000$}
 	\label{fig:echo1}
 \end{subfigure}
 \hspace{0.2in}
 \begin{subfigure}[b]{0.45\textwidth}
 	\includegraphics[width=1.1\textwidth]{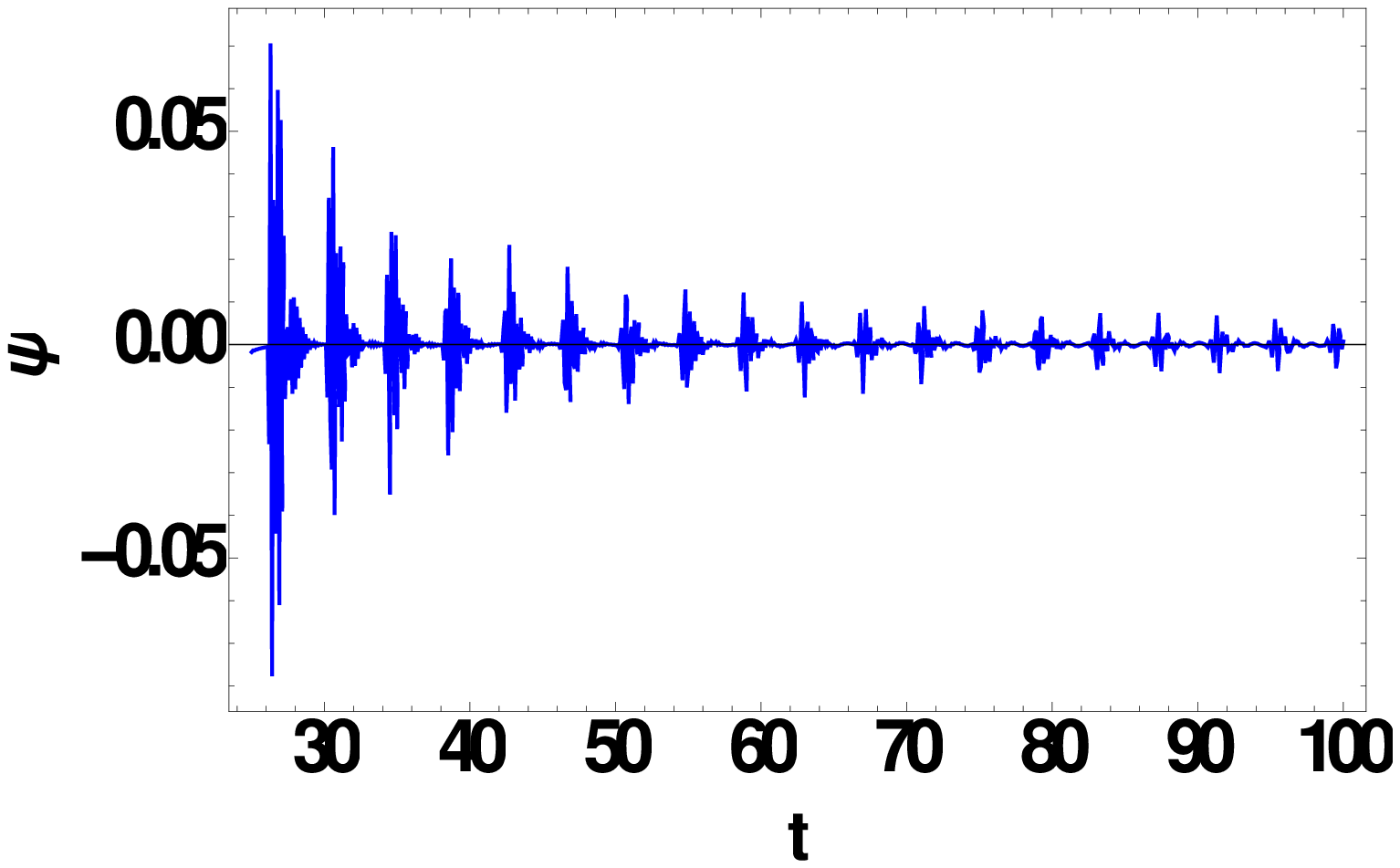}
 	\caption{ For $n=2600$}
 	\label{fig:echo2}
 	\end{subfigure}
 	\caption{Time domain profiles showing distinct echo pattern for very large $n$ geometries at $\ell=10$ and $m=1$. An initial Gaussian signal, given as $\psi(u,0) = e^{\frac{-(u-10)^2}{100}}$, is evolved over the u-v null grid as done in Sec.IV }
 \end{figure}

\noindent For our family of wormholes as we go to higher $n$ geometries, we get distinct echo patterns. With an increasing $n$, the peaks of the effective potential become sharper resembling delta-functions, hence resulting in sufficient reflection of waves to generate prominent echoes. The time domain profiles for two geometries with different $n$
values are shown in Fig.(\ref{fig:echo1}) and Fig.(\ref{fig:echo2}). Echoes are clearly
visible in these profiles. Note that there is no particular reason for
choosing $n=2000,2600$, except the fact that echoes are more prominent for large $n$.

 
 \noindent  For smaller $n$, echo patterns may be better observed with the help of a `cleaning' procedure of the time domain profile as described in \cite{echo7}. Ghersi et.al. \cite{echo7} have studied the scattering of wave packets from a Morris-Thorne \cite{mt} wormhole geometry which is constructed by joining two Schwarzschild geometries of equal mass at some $r_0 >  2M$. 
 The potential barrier for such a wormhole is symmetric about the throat hence forming a cavity and harbouring echoes in the time evolution of a wave packet. Since the amplitude of the echoes is very small in comparison to the entire spectrum, one needs to subtract the effect of scattering due to the black hole (i.e. a single potential barrier) in order to clearly observe the echoes. Through this process the authors in \cite{echo7}  `clean' the double barrier signal from the back-scattering occurring from the tail of the potential barrier, which, if not removed, suppresses the echo pattern. 
 
\noindent For scalar waves in our wormhole geometry we can perform a similar `cleaning' procedure too. In small `n' geometries, the potential barrier is smooth and hence has a wide tail which results in sufficient back scattering. Also the potential peaks do not have enough separation to provide strong reflections. So the echoes, even if they exist, are very weak and get damped easily. Thus, a similar `cleaning' procedure might help us in observing the presence of echoes for smaller `n' values which are normally not observable in the full spectrum. To proceed, we need a single barrier potential such that this single barrier, along with its mirror image, generates the double potential barrier. For this we consider only the peak present in the positive $\ell$ side. To create such a single barrier we have restricted the potential as given in eq.(\ref{eq:potential}) to $\ell \geq 0$. For $\ell <0$ we set the potential to a constant value 2 as the value of our double barrier potential around $\ell=0$ is always 2 for $m=1$. Hence, the `cleaned' profile will be given by: $\psi_{clean} = \psi_{double barrier} - \psi_{single barrier}$. This will leave us with echoes produced by the double barrier. Even though such a cleaning process is non-unique as we can always multiply some constant factor to the $\psi$ of the single barrier and still get the echo as a remnant of the subtraction process, it serves our purpose 
of demonstrating the existence of echoes for lower $n$ geometries.\\

 \begin{figure}[h]
 \centering
\begin{subfigure}[t]{0.45\textwidth}
  \centering
	\includegraphics[width=\textwidth]{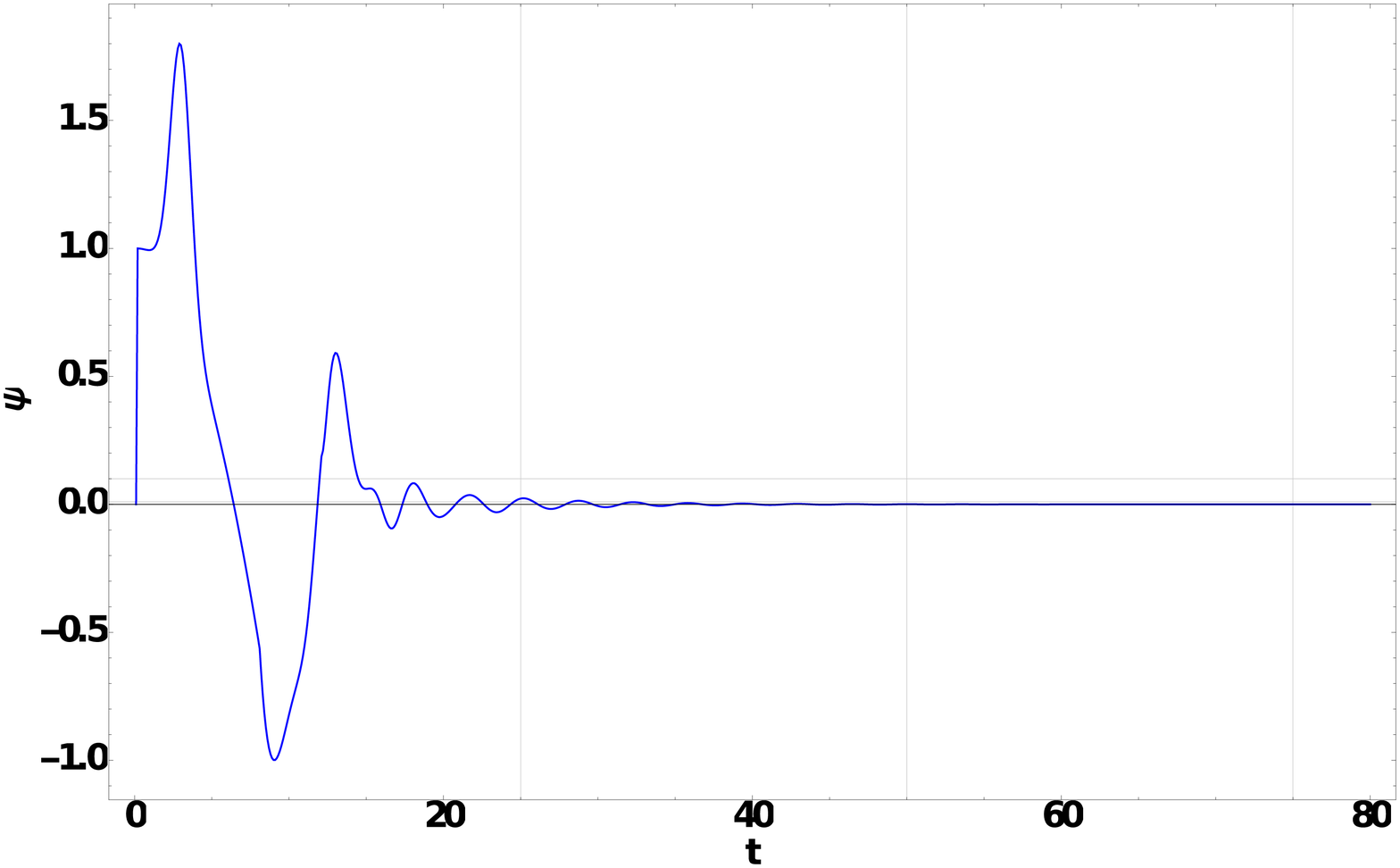}
 	\caption{ The `unclean' full spectrum for the double barrier.}
 	\label{fig:unclean_echo}
 \end{subfigure}
 \begin{subfigure}[t]{0.45\textwidth}
 	\centering
 	\includegraphics[width=\textwidth]{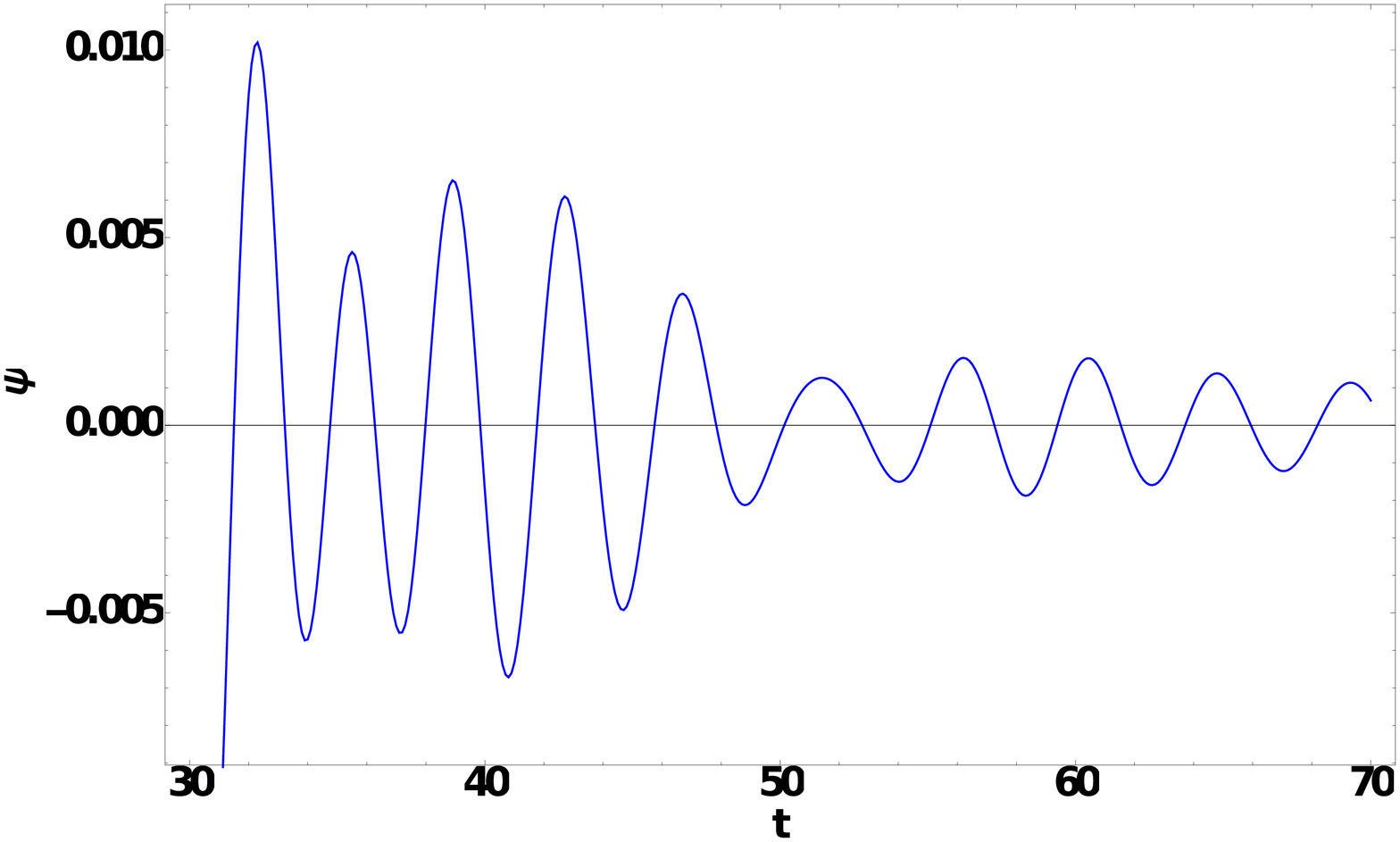}
 	\caption{Spectrum as obtained after `cleaning'.}
 	\label{fig:clean_echo}
 	\end{subfigure}
 	\caption{For the $n=100$ geometry the time evolution is observed at $\ell =5$ and $m=1$ with initial Gaussian impulse $\psi (u,0) = Exp(-\frac{(u-3)^2}{0.84})$. (a) Echoes are not
 	visible in the uncleaned spectrum. (b) Emergence of echoes 
 	after cleaning.}
 	\label{fig:echo3}
 \end{figure} 

\noindent The significance of the cleaning procedure is evident in Fig.(\ref{fig:echo3}) where we observe that for $n=100$ the echoes indeed exist and are observable after we clean the profile of the effect of the single barrier. We find that the echoes in Fig.(\ref{fig:clean_echo}) have an amplitude of the order of $10^{-2}$ lower than the original spectrum.

\noindent Thus the existence of echoes in a full spectrum without `cleaning' will be a tell-tale sign for a higher $n$ geometry. For high values of $n$, the wormhole (a 2D slice embedded in 3D Euclidean space) resembles two flat sheets with holes  which are connected by a cylinder. The wormhole throat is the radius of the cylinder which connects the flat regions. The effective potential
for scalar waves looks close to a double delta-function potential, for which
echoes are found. However, if we restrict ourselves to small $n$ geometries, as is done in the QNM studies here, we can safely comment that the echoes do not 
appear to have a dominant presence or contribution (and can be found only after
cleaning the spectrum). 

\section{\bf Summary and concluding remarks}

\noindent We summarize below the results obtained by us.

\noindent Working with a family of wormholes of which Ellis--Bronnikov spacetime 
is a special case, we first show how one may model the matter required for this generalised family. It turns out that it is possible to generate the members of this family of geometries, by using a massless phantom
scalar field and extra matter (absent for EB spacetime) satisfying the ANEC. 

\noindent Our next result is about the effective potentials for scalar perturbations.
We obtain the general conditions on the metric function for which the effective potential 
could be a double barrier. Thereafter, we illustrate the conditions using the family of wormholes dealt with in this paper. 

\noindent Subsequently, we focus on the scalar quasinormal modes for this family of Lorentzian wormholes. We obtain the QNM frequencies mainly using two methods: Prony fitting and direct integration and observe that both methods are equally suitable for our wormhole family.  

\noindent Having found the fundamental QNMs, we attempt to 
distinguish between the members in this wormhole family, through their values and 
dependencies on metric parameters and azimuthal number $m$.
The fact that the member wormholes of this family have different geometries
for different $n$ can also be visualised through their embedded shapes in a background Euclidean space. 

\noindent 
It turns that from the QNMs one may be able to distinguish between wormhole geometries, for 
lower $n$ values. For higher $n$, we cannot unfortunately make such a distinction just from
the QNMs, as the effective potentials become nearly identical, thereby making any such
attempt, largely difficult. This inability to distinguish between different, higher $n$ geometries
through QNMs, is not a shortcoming of the methods used for finding them but
is an innate feature of this wormhole family.
Note also that the QNMs corresponding to lower $m$ values are better suited for distinguishing between the geometries, as for higher $m$, the QNMs are similar in value
due to similar effective potentials.

\noindent Finally, as another way to distinguish different geometries, we have studied the occurrence of `echoes' through the time evolution of the field for different members (`$n$'
values) of the family. We find that although all $n>2$ geometries possess double barrier potentials, for low $m$ modes distinct echoes appear only at very large values of $n$. This happens because of the fact that for small $n$ wormholes the peaks of the potential are closely spaced and hence the reflections are not strong enough to produce 
clearly visible echoes. The echo signals
get damped giving way to the 
dominant QNMs of the wormhole. Using a `cleaning' procedure of subtracting the scattering of the single barrier from the full spectrum helps us in observing the echoes, if they exist, 
for such lower $n$ geometries too. On the other hand for large $n$, the peaks are well separated and we get distinct echoes, even without cleaning.

\noindent Thus, in summary, it is fair to say that the QNMs can be used as a distinguishing tool for small $n$ geometries while the presence of echoes in the (`uncleaned') time domain profile
is a clear indication of a large $n$ geometry.

\noindent We conclude with a remark. Our work, as reported here, provides a 
fairly thorough study of the geometry, matter, scalar
QNMs and echoes in this family of ultrastatic wormholes. This is a
first step in establishing this family of geometries as a viable template for
wormholes.
A useful future endeavour would be to analyse gravitational perturbations
which will enable us to use GW observations directly to test our
results and address the existence question of wormholes. We hope to pursue such 
investigations in future. 


\

\noindent {\bf Acknowledgments}

\noindent The authors thank Alok Laddha for his valuable comments. PDR thanks Indian Institute
of Technology, Kharagpur, India for support and for allowing her to use available facilities there.

\appendix
\section{Expressions of fitting functions for real and imaginary parts of $\omega_{QNM}$}

\noindent  In this brief appendix  we provide the details of the approximate analytical fit 
for the QNMs, as obtained using {\em Mathematica 10}. The Appendix is a supplement for the
Section IV C 1. 

\noindent When the real part of QNM frequency is fitted to the approximate analytical model given in eq.(\ref{eq:fit}) we get different values of the parameters, for different modes. Using the NonlinearModel fit package in {\em Mathematica} we do the fitting for two different values of $m$. For $m=1$, $b_0=1 $, $c=1$ we get,
\begin{equation}
\begin{split}
    \omega_r = 1\, +\, 0.6039890286502718\,n^{- 0.2684759097112008}\, + 0.03903126863165312\,n^2 \,-\\ 0.00820494335563206\,n^3 + 0.0006570415754729218\,n^4 - 0.00001867310871218107\,n^5
    \label{eq:fit_realm1}
\end{split}
\end{equation}
while for $m=10$, the model fitting gives,
\begin{equation}
\begin{split}
        \omega_r = 10\,+ 0.5455998550262489\,n^{0.5385664342040101}\,- 0.05290738182198302\,n^2\,+\\ 0.008299843261282835\,n^3\,- 0.0005644278402079367\,n^4 + 0.000014447119167250097\,n^5.
        \label{eq:fit_realm10}
\end{split}
\end{equation}
We have used machine precision (precision upto 16 decimal places) for fitting the frequencies and obtaining the fits in $Mathematica $ 10. 
Similarly, we can fit the imaginary part of the QNM frequency 
to the model given in eq.(\ref{eq:fit_im}) and obtain the parameters for $m=1$ as,
\begin{equation}
\begin{split}
    \omega_i = 0.6566372754281777\, n^{-0.1751876043748488}\,- 0.035808037683933445\,n^2\,+\\ 0.008127709544257126\,n^3\, -0.0006761084313479303\,n^4\,+ 0.000019652105439554295\,n^5 
\end{split}    
\end{equation}
while for $m=10$,
\begin{equation}
\begin{split}
    \omega_i = 0.1 \,( 16.39113462168537\, n^{- 1.7818703867902201}\,+ 0.019105687801418456\,n^2\,-\\ 0.003848586251352543\,n^3\,+ 0.00029441364674005576\,n^4\,- 8.037371021227578 \times 10^{-6} n^5).
\end{split}    
\end{equation}

\end{document}